\documentclass[%
 aip,
 amsmath,amssymb,
 preprint,%
]{revtex4-1}

\usepackage{graphicx}
\usepackage{dcolumn}
\usepackage{bm}

\usepackage[utf8]{inputenc}
\usepackage[T1]{fontenc}
\usepackage{mathptmx}

\begin{document}

\title[Singlet fission/halide perovskite interfaces]{A study of singlet fission-halide perovskite interfaces}

\author{A. R. Bowman}
\affiliation{ 
Cavendish Laboratory, Department of Physics, University of Cambridge, J.J. Thomson Avenue, Cambridge, CB3 0HE, U.K.}
\author{S. D. Stranks}
\affiliation{ 
Cavendish Laboratory, Department of Physics, University of Cambridge, J.J. Thomson Avenue, Cambridge, CB3 0HE, U.K.}
\affiliation{
Department of Chemical Engineering and Biotechnology, University of Cambridge, Philippa Fawcett Drive, Cambridge, CB3 0AS, U.K.}
\author{B. Monserrat}
 \email{bm418@cam.ac.uk}
\affiliation{ 
Cavendish Laboratory, Department of Physics, University of Cambridge, J.J. Thomson Avenue, Cambridge, CB3 0HE, U.K.}
\affiliation{Department of Materials Science \& Metallurgy, University of Cambridge, 27 Charles Babbage Road, Cambridge, CB3 0FS, U.K.}

\date{\today}

\begin{abstract}
A method for improving the efficiency of solar cells is combining a low-bandgap semiconductor with a singlet fission material (which converts one high energy singlet into two low energy triplets following photoexcitation). Here we present a study of the interface between singlet fission molecules and low-bandgap halide pervoskites. We briefly show a range of experiments screening for triplet transfer into a halide perovskite. However, in all cases triplet transfer was not observed. This motivated us to understand the halide perovskite/singlet fission interface better by carrying out first-principles calculations using tetracene and cesium lead iodide. We found that tetracene molecules/thin films preferentially orient themselves parallel to/perpendicular to the halide perovskite's surface, in a similar way to on other inorganic semiconductors. We present formation energies of all interfaces, which are significantly less favourable than for bulk tetracene, indicative of weak interaction at the interface. It was not possible to calculate excitonic states at the full interface due to computational limitations, so we instead present highly speculative toy interfaces between tetracene and a halide-perovskite-like structure. In these models we focus on replicating tetracene's electronic states correctly. We find that tetracene's singlet and triplet energies are comparable to that of bulk tetracene, and the triplet is strongly localised on a single tetracene molecule, even at an interface. Our work provides new understanding of the interface between tetracene and halide perovskites, explores the potential for modelling excitons at interfaces, and begins to explain the difficulties in extracting triplets directly into inorganic semiconductors.
\end{abstract}

\maketitle

\section{\label{sec:Introduction}Introduction}

Material surfaces and interfaces are important in a range of technologies and govern processes including sample growth, ion mixing, charge transfer, and electronic passivation\cite{Yadav2017,Martha2009,Yang2018a,AndajiGarmaroudi2020,Stadlober2006,DeQuilettes2016a,Zhang2015}. An interface of particular interest for the next generation of solar cells is that between a singlet fission material and an inorganic semiconductor. Singlet fission materials have the unusual property that when a singlet is generated (following photoexcitation) it is rapidly converted into two triplets~\cite{Smith2010}. If combined with an inorganic semiconductor harvesting low energy wavelengths of the solar spectrum, such a solar cell could surpass the Shockley-Queisser single junction efficiency limit of 33 \%~\cite{Tayebjee2015}. Key to this technology working is the ability to extract at least part of the triplet excitons from the singlet fission material. Ideally triplets would be transferred directly into, or separated at, an interface with an inorganic semiconductor (exciton transfer and exciton dissociation). Triplet excitons cannot undergo F\"{o}rster energy transfer~\cite{Scholes2003}, as they do not have a dipole moment (unlike singlet excitons). Therefore, charge transfer must proceed via Dexter processes, which are far shorter-ranged in nature and governed by wavefunction overlap~\cite{Dexter1953}. 

In studies of clean silicon surfaces with singlet fission materials deposited on them, the component of triplet transfer has been negligible~\cite{Piland2014,Macqueen2018}. However, extraction of triplets has been achieved into PbS and PbSe quantum dots, as well as more recently silicon~\cite{Tabachnyk2014,Thompson2014,Einzinger2019}. In these cases there has been some modification of the inorganic semiconductor's surface or a direct chemical bond from it to the singlet fission material. More generally it has proven difficult to extract triplets directly from singlet fission materials into inorganic semiconductors, and a full understanding of the difficulties involved is lacking. 

Here, we explore the interface between singlet fission materials and halide perovskites. First, we briefly present experiments screening for triplet transfer from singlet fission materials to low-bandgap (< $1.25$ eV) halide perovskites. Despite modified fission activity in the organic, in all cases we observed singlet transfer or no transfer to the halide perovskite. These experiments motivated us to understand the interface better via a first principles computational study. We modelled the interface between the singlet fission material tetracene and caesium lead iodide (CsPbI$_{3}$, noting that in our computations the bulk halide perovskite has a bandgap lower than tetracene's triplet energy)~\cite{Green2014}. We find that the tetracene/CsPbI$_{3}$ interface is not very energetically favourable to form and films of tetracene form with the long axis of the molecule perpendicular to the halide perovskite surface. We also present highly speculative calculations suggesting that triplets remain strongly localised on tetracene, even at the interface. Our computational results go some way to explaining experimental observations and suggest a clear route forward for future studies.

\section{Brief experimental screening}

We fabricated approximately 150 samples as low-bandgap (< $1.25$ eV) halide perovskites/singlet fission bilayers (see schematic, supplementary information Figure~\ref{fig:Schem_and_AFM}a). A full list of all processing routes explored can be found in SI~\ref{Screening_list}. 

We probed triplet transfer by applying a magnetic field to the singlet fission/halide perovskite bilayer and observing the change in the halide perovskite's photoluminescence (PL) and via time correlated single photon counting (TCSPC). If there is net triplet transfer or dissociation at the interface the halide perovskite's PL will reduce at high magnetic field as it will receive fewer triplets~\cite{Merrifield1968}. 

Two singlet fission materials were used in our screening experiments: tetracene and 1,6-Diphenyl-1,3,5-hexatriene (DPH)~\cite{Merrifield1969,Wilson2013,Dillon2013b,Wakasa2015}. The triplet energies in these molecules ($\sim 1.3$ eV and $1.5$ eV respectively) are larger than the selected halide perovskite's bandgap ($\sim 1.25$ eV) and they both have high triplet yields (with > $50$ \% of singlets converted to triplets). 

When we began this study, it had recently been observed that adding small quantities of transition metals into halide perovskites could change their work function~\cite{Klug2017}. A similar effect could be seen by changing the quantity of caesium at the A site of the halide perovskite~\cite{Prasanna2017}. In both cases the halide perovskite's bandgap remains constant. Halide perovskites with different work functions were fabricated according to these two methods. All samples had either 50:50 or 25:75 ratios of lead to tin, corresponding to bandgaps in the $1.2$ - $1.25$ eV range. We fabricated films that were thin ($50$ - $200$ nm) to increase any singlet fission contribution to the halide perovskite's photoluminescence. 

We present an example of the results obtained from these screening experiments in supplementary Figure~\ref{fig:Tetracene_ex_results}, for a tetracene/FA$_{0.9}$Cs$_{0.1}$Pb$_{0.25}$Sn$_{0.75}$I$_{3}$ bilayer (with tetracene evaporated on to the halide perovskite, where FA is formamidinium). We present the magnetic PL response of evaporated tetracene in Figure~\ref{fig:Tetracene_ex_results}a, showing an increase in PL at high magnetic field, as expected. In Figure~\ref{fig:Tetracene_ex_results}b the photoluminescence of the bilayer is presented (from $500$ nm to $1000$ nm). Tetracene can be weakly observed in the $500$ nm to $700$ nm region (it is weak due to the camera used, see figure caption), and the halide perovskite PL peaks at $\sim 975$ nm. We were able to selectively monitor the halide perovskite's PL change with magnetic field by using a $900$ nm long-pass filter, as shown in Figure~\ref{fig:Tetracene_ex_results}c. At low magnetic field the change in PL is < $0.2$ \%, which is within the experimental noise. However, at high magnetic field the halide perovskite's PL increases to a significant level (> $0.4$ \%), indicating net singlet transfer from tetracene. Singlet transfer is confirmed by TCSPC, as shown in Figure~\ref{fig:Tetracene_ex_results}d. By exciting the bilayer above and below tetracene's bandgap and observing the halide perovskite's time resolved photoluminescence (TRPL), we observe the longer lived component of this TRPL coincides well with the TRPL from tetracene (in the bilayer). Furthermore, we find the halide perovskite's TRPL shows nothing longer lived than the singlet transfer, suggesting any triplet transfer from tetracene is negligible (noting triplets are typically  longer lived than singlets). 

In all cases explored, we observed singlet fission in the organic layer but no net triplet transfer to the halide perovskite (though singlet transfer was readily observed). To better understand organic/inorganic interfaces, and potentially explore the reasons for the lack of triplet transfer from singlet fission materials, we modelled a proto-typical singlet fission/halide perovskite bilayer.

\section{Modelling background}

It has recently become possible to study some simple interfaces with first-principles computational methods. For example, density functional theory (DFT) calculations (corroborated by X-ray diffraction experiments) have helped confirm the geometry of the interface between tetracene and silicon~\cite{Janke2020,Niederhausen2020}, where tetracene thin films were found to orient with their long axis perpendicular to the semiconductor surface. 

Here, we model tetracene and CsPbI$_{3}$ using DFT. We selected CsPbI$_{3}$ for this study as the cation has spherical symmetry (unlike methylammonium or formamidinium) and an inorganic lead halide perovskite reduces the number of atomic species in the system, simplifying modelling while maintaining the same basic electronic structure. We note that the electronic structure of halide perovskites is governed by the metal/halide framework, so as long as this is maintained we anticipate results will be comparable for different compositions. As the literature has shown that it has been difficult to obtain triplet transfer to several different semiconductors, we postulate that the reason for lack of triplet transfer lies primarily within the singlet fission material. This is further supported by halide perovskites sensitising triplet states in singlet fission materials~\cite{Lu2019}. Therefore, our modelling focuses on correctly reproducing tetracene's electronic states (at the expense of correctly modelling the halide perovskite's electronic states). 

We first discuss each system in isolation and quantify the effects of surface terminations and bandgap corrections. We then present geometry arrangements and resulting electronic structures of a single tetracene molecule and bulk tetracene films on a halide perovskite surface. As in previous studies~\cite{Janke2020}, we found post-DFT $GW$ and BSE calculations (required to calculate excitonic states) of full interfaces were not feasible. Instead, we present highly speculative toy models between tetracene and a halide perovskite-like structure. We focus on correctly reproducing tetracene's electronic states at the interface (at the expense of the halide perovskite). All computational details for subsequent sections can be found in supplementary information~\ref{sec:Comp_details}.

\section{\label{sec:Tetracene_alone}Tetracene}

Tetracene is one of the most widely studied singlet fission materials\cite{Merrifield1969,Avakian1968,Smith2010,Tayebjee2013,Stern2017a}. It undergoes singlet fission endothermically and, consequently, the process proceeds at a slower pace than in other molecules\cite{Wilson2013}. Amongst singlet fission materials, its triplet energy is one of the highest, making it one of the most relevant for solar energy applications\cite{Wu2014}. Initially, we carried out geometry relaxations of bulk tetracene unit cells using several different exchange correlation functionals and Van der Waals corrections. We found the PBE generalised gradient approximation, coupled to a Tkatchenko-Scheffler (TS) Van der Waals semi-empirical correction, best reproduce experimental tetracene lattice parameters, in agreement with other analyses~\cite{Tkatchenko2009,Janke2020} (see supplementary information Table~\ref{tab:lattreslts}). We used this functional and Van der Waals correction in all subsequent calculations.

\begin{figure}
    \includegraphics[width=0.5\textwidth]{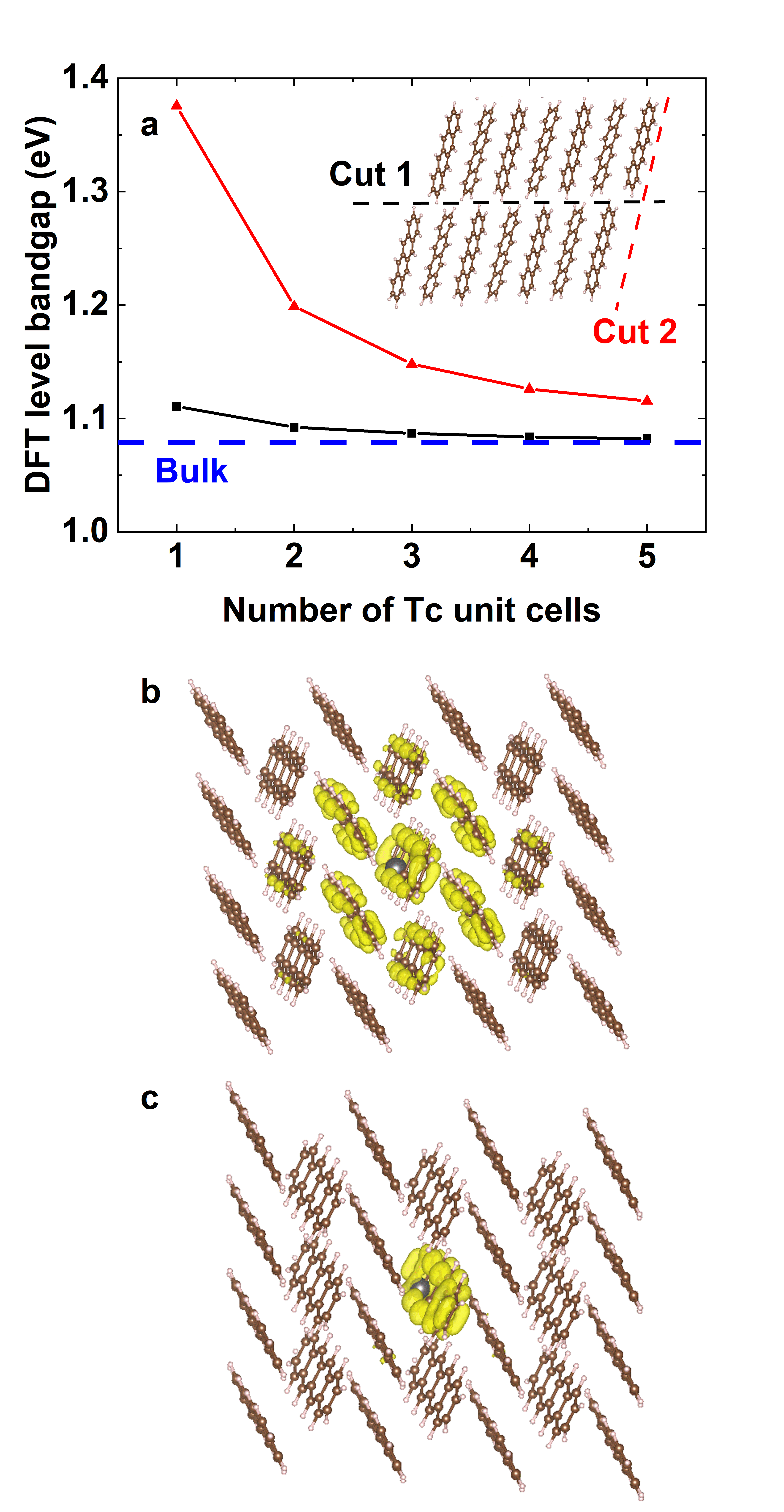}
    \caption{\label{fig:Tc_fig}  a) The DFT-level bandgap of relaxed tetracene films with different surface terminations versus the number of tetracene molecules in the vacuum direction. Inset shows the two surface terminations considered and dashed blue line the bulk result. The singlet and triplet electron charge densities, for hole (grey sphere) fixed on a carbon atom are presented in b) and c) respectively.}
\end{figure}

In Figure~\ref{fig:Tc_fig}a we plot the DFT-level bandgap for different relaxed tetracene surfaces versus the number of tetracene repeating units in the non-periodic direction. Modelled surfaces are termed `cut 1' and `cut 2' (other cuts were not commensurate with the halide perovskite unit cell), as shown on the inset. While cut 1 has relatively little effect on the DFT-level bandgap, cut 2 results in a bandgap increase of $\sim$ $0.25$ eV for a single repeating unit, which reduces to within $0.1$ eV by three repeating units. This demonstrates there is relatively little electronic interaction between tetracene layers (i.e. in the direction perpendicular to cut 1); rather, electronic states are mostly localised to one layer.

Tetracene's (optically) excited states typically exist as singlets and triplets. We used a one-shot $G_{0}W_{0}$ calculation together with the Bethe-Salpeter equation to calculate the singlet and triplet states of bulk tetracene. The lowest energy singlet and triplet states of relaxed tetracene have energies of 2.08 eV and 1.24 eV respectively, which increase to 2.22 eV and 1.27 eV when using experimental lattice parameters, in good agreement with experimental results~\cite{Tomkiewicz1971,Wilson2013}. While it is not possible to plot excitonic wavefunctions (as these are two-particle states requiring six spatial coordinates), it is possible to plot electron or hole charge densities with the other particle fixed in place. In Figure~\ref{fig:Tc_fig}b and c we plot the electronic charge density for the lowest energy singlet and triplet states respectively, with the hole fixed at a carbon atom (grey sphere). The singlet state is delocalised over several molecules but the triplet is almost fully localised to a single molecule, as has been previously discussed by others~\cite{Alvertis2020}. We found this to be the case for the hole being fixed at several different positions within the tetracene film. Furthermore, neither the singlet nor triplet states have significant wavefunction overlap between planes of the film (in direction perpendicular to cut 1), again showing that films have quasi-two-dimensional electronic states.

\section{Caesium lead iodide}
\label{sec:halide perovskite_alone}

\begin{figure}
    \includegraphics[width=0.7\textwidth]{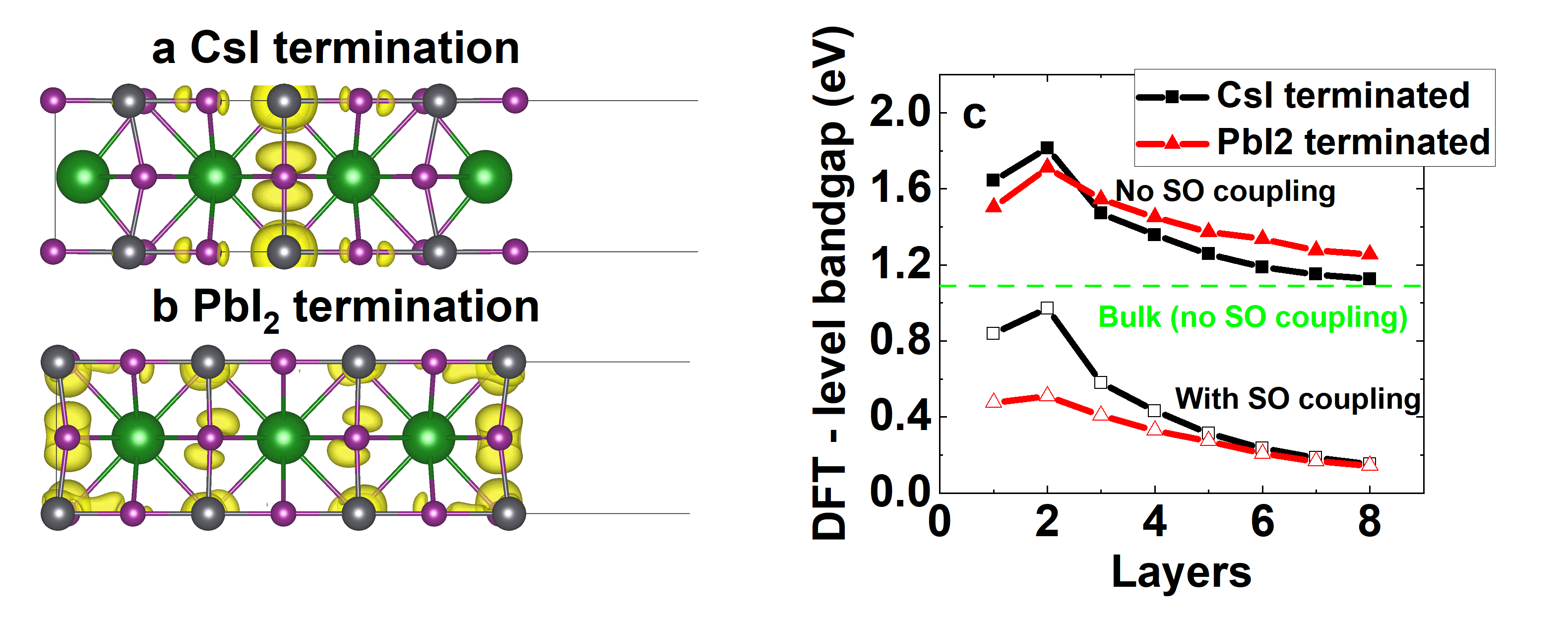}
    \caption{\label{fig:Perov_fig} a) and b) plot the valence band charge density (at the band edge) of CsI and PbI$_{2}$ terminated surfaces. In c) the DFT-level bandgap is presented for different numbers of repeating units in the vacuum direction, for (100) CsI and PbI$_{2}$ terminated surfaces, with and without spin-orbit (SO) coupling. The dashed line marks the bulk result in the case of no spin-orbit coupling. In a)-b) all calculations include spin-orbit coupling.}
\end{figure}

We discuss the electronic structure of bulk CsPbI$_{3}$ in supplementary information~\ref{App:Bulk_perov}. When we consider an interface there are many possible surface terminations for a cubic inorganic structure. First-principles calculations on related halide perovskites suggest (100) surfaces are the most likely to form so our study focsues on these surfaces~\cite{Haruyama2014}. We consider two possible surface terminations of this plane: PbI$_{2}$ and CsI. In all our modelling the same surface was used on both sides of the halide perovskite, mitigating surface dipole effects. We present the valence band charge densities for these two relaxed surface terminations in Figure~\ref{fig:Perov_fig}a and b. While the CsI termination has the valence band confined to the bulk, PbI$_{2}$ termination has a significant surface contribution (i.e. a dangling bond). This gives two significantly different situations, so our subsequent calculations explore both surfaces. We note that PbI$_{2}$ surfaces are lower in energy so are more likely to form in reality. In both materials the conduction band is confined to the bulk. 

The electronic structure of halide perovskites is fundamentally three-dimensional, so it is strongly affected by quantum confinement. In Figure~\ref{fig:Perov_fig}c we present the DFT-level bandgap of CsPbI$_{3}$ for both relaxed surface terminations, as a function of the number of repeating units in the vacuum direction, both with and without spin-orbit coupling. In all cases a significant increase in bandgaps is found when compared to the bulk, as expected. We find CsI terminated surfaces have slightly larger bandgaps, which is attributed to more confined states (c.f. Figure~\ref{fig:Perov_fig}a and b).

\section{Tetracene molecules on halide perovskite surfaces}
\label{sec:Molec_int}

\begin{figure}
    \includegraphics[width=0.5\textwidth]{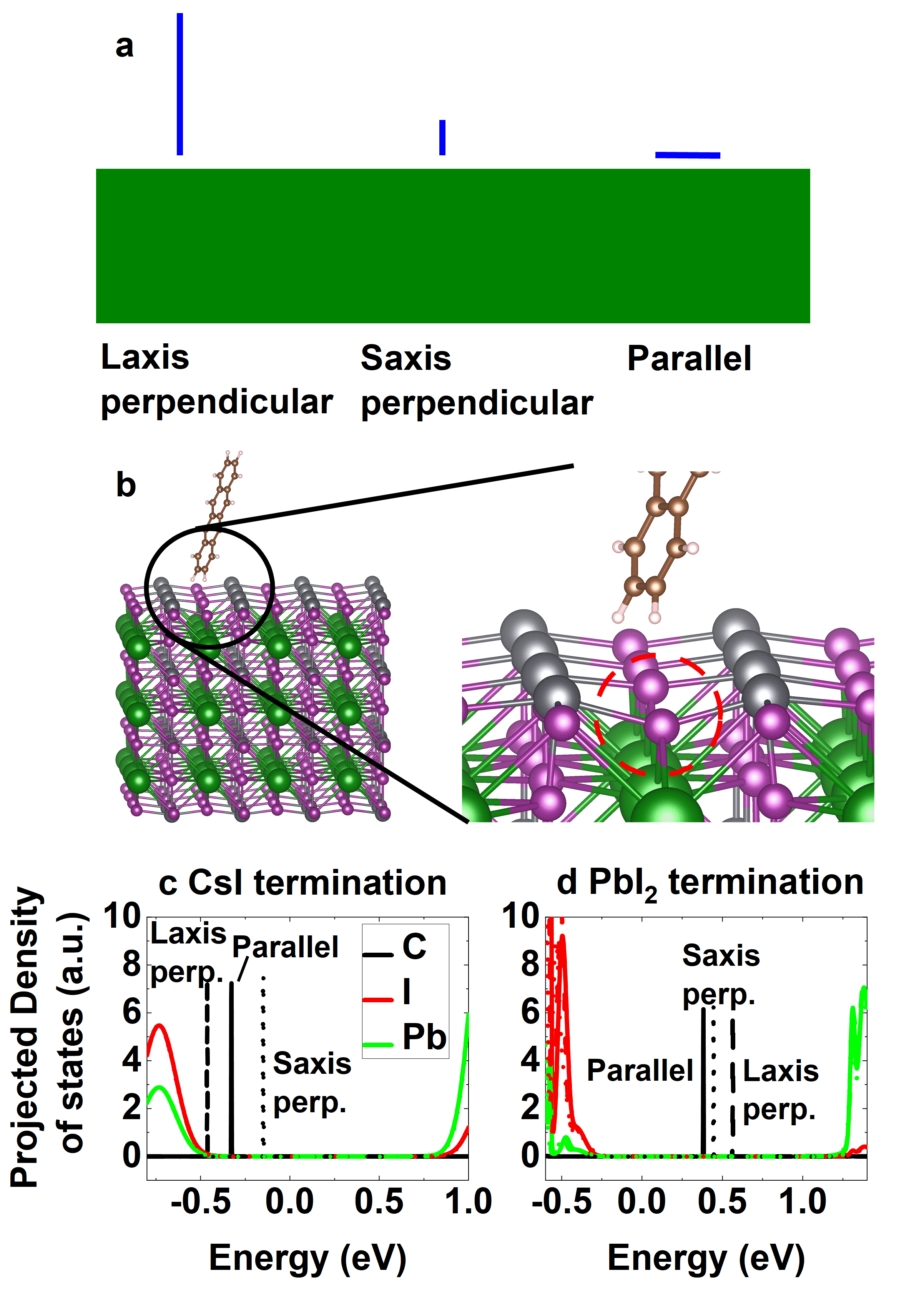}
    \caption{\label{fig:Tc_molec} The three possible tetracene orientations (blue lines) on the halide perovskite surface (green slab) are presented in a). An example of a relaxed geometry is shown in b), for PbI$_{2}$ surface termination with tetracene's long axis perpendicular to the surface. The region of interaction is enlarged, showing the halide perovskite surface distortion due to the tetracene. c) and d) show the projected density of states, in arbitrary units, without spin-orbit coupling, for the three tetracene orientations for CsI and PbI$_{2}$ terminated surfaces respectively. Energy has been shifted arbitrarily to align the halide perovskite valence bands. In all cases the carbon states are the highest occupied states.}
\end{figure}

To simulate a halide perovskite thin film, we modelled three halide perovskite layers in the non-periodic direction, as inter-atomic distances in the centre of this structure were comparable to distances in bulk CsPbI$_{3}$ (see supplementary information Table~\ref{tab:Perov_lat}) and computation was not too expensive. As has been discussed elsewhere~\cite{Niederhausen2020,Hlawacek2013,Janke2020}, there are three main orientations for a tetracene molecule to lie on a halide perovskite surface: with the long axis of the molecule perpendicular; the short axis of the molecule perpendicular; or face-on to the halide perovskite (termed parallel, see Figure~\ref{fig:Tc_molec}a). 

We carried out geometry relaxations for the three tetracene orientations on both halide perovskite surfaces (i.e. six simulations). The relaxed energy is relatively independent of where the tetracene is initially placed above a semiconductor~\cite{Janke2020}, so we only considered one starting position for each geometry relaxation. We present an example of a relaxed geometry in Figure~\ref{fig:Tc_molec}b. The tetracene molecule is observed to remain unchanged, while the halide perovskite surface distorts. We found qualitatively similar results in all simulations. This is reasonable as halide perovskites have surprisingly low Young's moduli for inorganic semiconductors, comparable to those of relevant organic crystals~\cite{Sun2015,Jhou2019,Wortman1965,Chen2016}.

The formation energy between the halide perovskite surface and the tetracene molecule is given by E$_{formation}$=E$_{interface}$-(E$_{perov,slab}$+E$_{tetracene,molecule}$), that is the energy of the interface less the energy of the halide perovskite slab and tetracene molecule in vacuum.  We present E$_{formation}$ for the six geometries considered in Table~\ref{tab:TcmolecE}. Results suggest the parallel tetracene arrangement has the most favourable formation energy. This is in agreement with other examples of rod-like molecules on the surface of inorganic semiconductors~\cite{Janke2020,Niederhausen2020,Hlawacek2013}. Furthermore, in all cases we find the interaction energy is observed to be weak: for comparison the formation energy for a tetracene film is $\sim -2$ eV per molecule.

\begin{table}[h!]
	\centering
	\begin{tabular}{ m{4cm} | >{\centering\arraybackslash}m{3cm} }
		Tetracene orientation & $E_{formation}$ (eV)\\
		\hline
		\hline
		\textit{CsI termination} & \\
		Long axis perpendicular & $-0.50$  \\
		Short axis perpendicular & $-0.26$  \\
		Parallel & $-0.69$  \\
		\hline
		\textit{PbI$_{2}$ termination} & \\
		Long axis perpendicular & $-1.03$ \\
		Short axis perpendicular & $-0.82$ \\
		Parallel & $-1.25$  \\
	\end{tabular}
	\caption{\label{tab:TcmolecE}Formation energy, E$_{formation}$, for three orientations of a tetracene molecule on CsI and PbI$_{2}$ terminated halide perovskite surfaces.}
\end{table}

In order to understand band-alignment at each interface, we calculated the projected density of states (PDOS), with projection onto each species' atomic orbitals. We neglected spin-orbit coupling as this gives a similar halide perovskite bandgap to that obtained with spin-orbit coupling and a $G_{0}W_{0}$ correction at significantly less computational cost (see supplementary information section~\ref{sec:halide perovskite_alone}). Our results are presented in Figure~\ref{fig:Tc_molec}c and d. For CsI terminated surfaces, as the halide perovskite's valence band is localised in the bulk of the structure (c.f. Figure~\ref{fig:Perov_fig}a), its PDOS is unchanged for different tetracene orientations (Figure~\ref{fig:Tc_molec}c). For the (energetically favoured) parallel and short-axis perpendicular orientations, tetracene's valence band lies above that of the halide perovskite while for the long-axis perpendicular model its valence band aligns with the halide perovskite's valence band maximum. For PbI$_{2}$ termination, as the halide perovskite's valence band has some surface character, we find there is a small variation in the halide perovskite's PDOS for the different simulations (Figure~\ref{fig:Tc_molec}d). Importantly, tetracene's valence band is above the halide perovskite's in all cases, and significantly more offset from the halide perovskite's valence band than in the CsI terminated case. We attribute this to PbI$_{2}$ terminated surfaces being lower in energy than CsI terminated surfaces rather than a significant change in tetracene's electronic structure. Tetracene's valence band being above the halide perovskite's is indicative of tetracene being a good hole extracting material, as has been observed experimentally in halide perovskite/tetracene interfaces~\cite{Abdi-Jalebi2019}. 

\section{\label{sec:Full_int}Tetracene-halide perovskite thin film interface}

To model interfaces between bulk materials, unit cells with commensurate in-plane lattice parameters need to be found. We identified two tetracene/halide perovskite interfaces with tetracene's `cut 1' surface termination and one with `cut 2' surface termination. They are shown in supplementary information Figure~\ref{fig:Commens_uc}. We were not able to find any commensurate unit cells of reasonable size for tetracene's third possible exposed plane. These three simulations are termed `cut 1 no rotation', `cut 1 with rotation' and `cut 2'. The two cut 1 simulations have normal strains of < 3 \% and shear strain of $\sim 10$ \%, while cut 2 has both normal and shear strains on the order of $5$ \% (see supplementary information Table~\ref{tab:Comms_strain} for all strains). Two tetracene unit cells were used in the non-periodic direction for cut 1 models, and four tetracene units cells for cut 2 (based on results shown in Figure~\ref{fig:Tc_fig}a). This resulted in large cells for DFT simulations so no vacuum layer was used i.e. there were two interfaces between tetracene and the halide perovskite. This also reduced the possibility of any dipole effects. The distance between adjacent halide perovskite and adjacent tetracene layers was at least as large as the vacuum spacing required to prevent interaction between adjacent layers of the same material in the non-periodic direction. As tetracene and halide perovskites have comparable Young's moduli, we allowed lattice parameters to vary in geometry relaxations~\cite{Sun2015,Jhou2019}. 

\begin{figure*}
\includegraphics[width=0.8\textwidth]{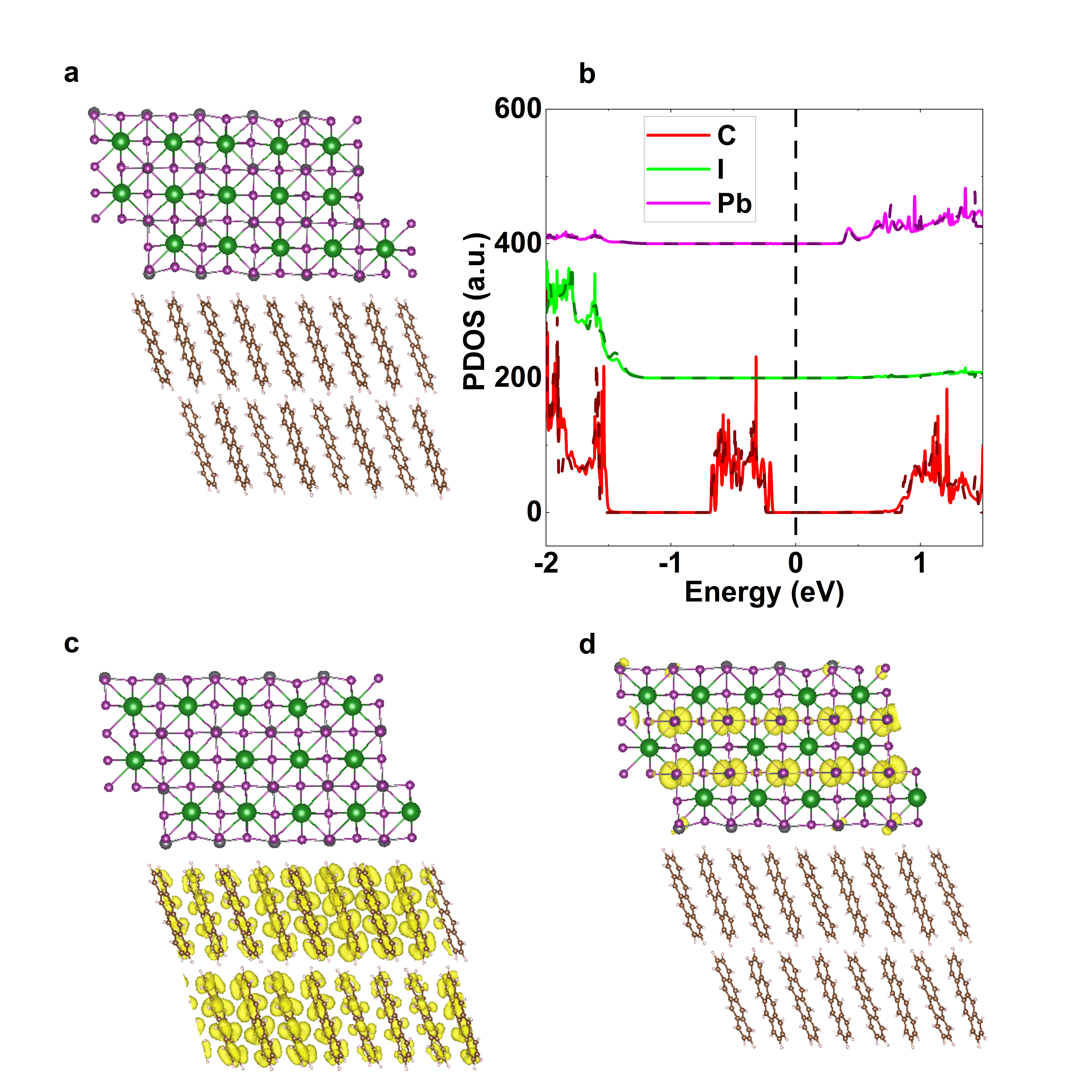}
\caption{\label{fig:Full_int} a) The fully relaxed geometry of cut 1 no rotation with PbI$_{2}$ surface termination. The corresponding projected density of states (PDOS) without spin-orbit coupling is shown in b). Here the dashed vertical black line marks the Fermi level, and dashed coloured lines mark the PDOS for isolated relaxed PbI$_{2}$ and tetracene slabs in vacuum. The valence and conduction band charge densities of this interface at band edges are presented in c) and d).}
\end{figure*}

We ran geometry relaxations for all simulation cells. All cuts ran successfully except cut 2 with CsI termination, where the halide perovskite structure fell apart during computation. We attribute this to the unit cell being too strained for the optimisation to complete and this geometry is not discussed further. We present a relaxed geometry in Figure~\ref{fig:Full_int}a for PbI$_{2}$ terminated cut 1 no rotation. In all cut 1 configurations, including that presented here, we found tetracene molecules to be almost entirely unchanged in position from a thin film of tetracene. For cut 2, we found the spacing between tetracene molecules reduced slightly (< $2$ \% change) in the non-periodic direction, especially for molecules adjacent to the interface. In all cases the halide perovskite surface distorted more significantly.

We define the interface formation energy per tetracene molecule as

\begin{equation}
	2\times \text{E}_{formation,pm}=\frac{\text{E}_{interface}-\text{E}_{perov,slab}}{N_{Tc}}-\text{E}_{tetracene,molecule}, 
\end{equation}
where $N_{Tc}$ is the number of tetracene molecules in the interface simulation. The factor of two arises from the double interface. We present formation energies in Table~\ref{tab:FullfilmsE}. Larger energy reductions are observed for CsI termination, which is explained by noting that bare CsI surfaces are less stable (higher in energy) than the PbI$_{2}$ surfaces, so bulk films provide more significant stabilisation. In both cases the two cut 1 geometries have almost identical energies, suggesting that tetracene's orientation on the halide perovskite surface is only weakly dependant on the halide perovskite. This is indicative of weak interaction between the two materials. For PbI$_{2}$ termination, we find cut 1 orientations are more favourable than cut 2, as is again expected for rod-like molecules~\cite{Hlawacek2013}. All formation energies are significantly higher than for bulk tetracene ($\sim -2$ eV per molecule) which suggests tetracene will not bind strongly to the halide perovskite surface. It may instead form pillars or other structures with only small contact regions with the halide perovskite, as seen in our experimental section (c.f. supplementary information Figure~\ref{fig:Schem_and_AFM}b) and has been discussed by others~\cite{Hlawacek2013}. 

\begin{table}[h!]
	\centering
	\begin{tabular}{l | >{\centering\arraybackslash}m{3.5cm} }
		Tetracene orientation & $E_{formation,pm}$ (eV) \\
		\hline
		\hline
		\textit{CsI termination} & \\
		Cut 1 no rotation & $-1.42$  \\
		Cut 1 with rotation & $-1.41$  \\
		\hline
		\textit{PbI$_{2}$ termination} & \\
		Cut 1 no rotation & $-0.95$  \\
		Cut 1 with rotation & $-0.95$ \\
		Cut 2 & $-0.92$ \\
	\end{tabular}
	\caption{\label{tab:FullfilmsE}Formation energies per molecule, $E_{formation,pm}$, for relaxed thin film interfaces.}
\end{table}

We show the projected density of states for PbI$_{2}$ terminated cut 1 no rotation in Figure~\ref{fig:Full_int}b (PDOS for other cut 1 simulations are presented in supplementary information Figure~\ref{fig:OtherPDOS}). As in the case of a single molecule, tetracene's valence band is located above the halide perovskite's. This was the case for all other simulations. Overlaid on this plot in dashed lines are the PDOS for halide perovskite and tetracene slabs (independently) isolated in a vacuum. For the halide perovskite, the isolated PDOS overlaps almost perfectly with the valence and conduction bands. We find tetracene's PDOS at the interface is slightly broadened with respect to isolated tetracene, which we attribute to each tetracene molecule being in a slightly different electronic environment at the interface. This PDOS demonstrates there is little interaction between the halide perovskite and tetracene, as their states can be well reproduced by isolated slabs. 

We present the valence and conduction band charge densities (at the band edge) for cut 1 no rotation, for PbI$_{2}$ termination, in Figure~\ref{fig:Full_int}c and d respectively. Both charge densities are isolated to one material only (tetracene and halide perovskite respectively), again implying that there is little change in electronic states at the interface. When looking further into the valence band (at the k-point corresponding to the valence band maximum), no states are found are found to have significant charge density in both the halide perovskite and the tetracene. 

\section{\label{sec:Small_int}Toy interface with exciton visualisation}

To model triplet excitons it is necessary to solve the Bethe-Salpeter equation. We attempted this on the full interfaces presented in the previous section, but found it was not computationally feasible. A similar limitation was recently highlighted in Janke and co-workers' study of pentacene and tetracene on passivated silicon surfaces~\cite{Janke2020}. Other studies have focused on the development of fragment-based (post-DFT) $GW$ and BSE calculations, allowing for modelling of interfaces~\cite{Bethe2020}. However, in fragment-based approaches the exchange interaction needs to be neglected, meaning singlet and triplet states cannot be differentiated.

Despite these limitations, we decided to carry out highly speculative calculations to explore what excitonic properties can be modelled at an interface with modern computational methods and resources. We note it is unclear whether full excitonic states will form at the interface, or whether charges dissociate prior to reaching the interface, but here we aim to increase understanding of the possible states at the interface and ascertain what current computational methods reveal. To this end we constructed small toy interfaces consisting of a single tetracene unit cell and a single (in plane) CsPbI$_{3}$ unit cell. In these models we oriented tetracene perpendicular to the interface (cut 1 orientation, as this was the most energetically favourable) and the halide perovskite again had three repeating units in the non-periodic direction. As we are interested primarily in tetracene's electronic states, the model's in plane lattice parameters were constrained to those of tetracene. To achieve this, one of the halide perovskite's in plane lattice parameters was increased by $\sim$ 20 \% and the other reduced by $\sim$ 3 \%. Again we modelled both CsI and PbI$_{2}$ terminations. We carried out geometry optimisations (with in plane lattice parameters constrained) without any vacuum spacing. However, for $G_{0}W_{0}$ and BSE calculations the lack of vacuum spacing resulted in interaction between adjacent unit cells in the non-periodic direction. Therefore, we introduced a vacuum layer (the same size as the `filled' unit cell) and used a Coulomb cutoff to prevent long-range interactions between repeating unit cells.

We calculated the PDOS of these small models (with vacuum spacing) and our results are presented in supplementary information Figure~\ref{fig:ToyPDOS}a and b. Electronic states were positioned similarly to those presented in previous sections, with the same atomic orbitals contributing to the halide perovskite's valence and conduction bands, and tetracene's valence band being above that of the halide perovskite. Importantly, we found the halide perovskite's bandgap had increased, due to the larger in plane lattice parameter increasing quantum confinement. Density of states calculations including spin-orbit coupling, plotted in supplementary information Figure~\ref{fig:ToyPDOS}c and d, demonstrate that for PbI$_{2}$ termination there is no bandgap in this system, preventing post-DFT calculations with spin-orbit coupling on this system.

We were able to carry out $G_{0}W_{0}$ and BSE calculations on these toy interface models. However, further simplifications were needed for calculations to proceed: we neglected the non-local commutator and we reduced the maximum reciprocal lattice vector size with respect to DFT calculations. The latter approximation is equivalent to reducing the cutoff energy in a DFT calculation, but was found to have only small effect (< $0.01$ eV) due to smaller reciprocal lattice grids being required in $G_{0}W_{0}$ calculations (see supplementary information~\ref{sec:Comp_details}). We found both approximations were reasonable by carrying out one calculation without these approximations - minimal difference in results was observed. 

$G_{0}W_{0}$ calculations on the toy interfaces, alongside on just tetracene and toy halide perovskite, allow for an estimation of the changes that would occur to the PDOS calculations already presented. Specifically, we present $G_{0}W_{0}$ corrections in supplementary information Table~\ref{tab:G0W0corrections}, focusing on the change in energy offset between halide perovskite and tetracene valence bands. In general our results point towards the difference in energy between the halide perovskite and tetracene valence bands being larger than what was calculated at DFT-level by $\sim$ 1 eV. As was discussed in section~\ref{sec:halide perovskite_alone}, this would be partly offset by calculations including spin-orbit coupling in the halide perovskite (reducing energy difference between the halide perovskite and tetracene valence bands), so we consider that the PDOS presented above give the correct qualitative conclusions. 

\begin{figure}[h!]
	\centering
	\includegraphics[width=0.8\textwidth]{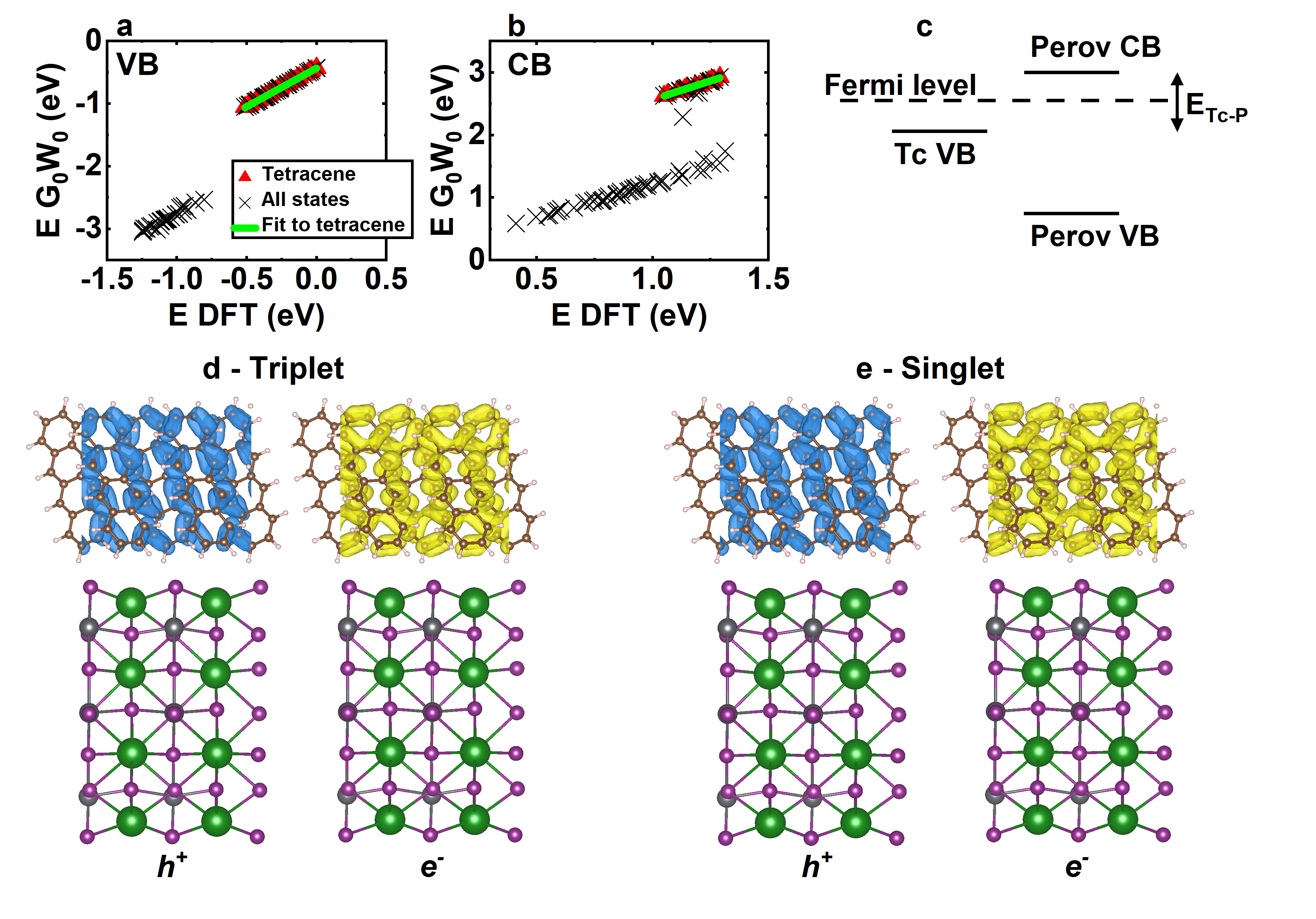}
	\caption{DFT energies versus $G_{0}W_{0}$ energies for the CsI terminated toy model's valence band (VB) and conduction band (CB), with spin-orbit coupling, are shown in a) and b). Legend in a) applies to a) and b). Schematic in c) shows relative band energies at the interface, with $E_{Tc-P}$ being the difference in energy between tetracene's valence band and the halide perovskite's conduction band. Average hole (h$^{+}$) and electron (e$^{-}$) charge densities for the lowest energy triplet and singlet excitons for the CsI terminated surface are plotted in d) and e), for the $G_{0}W_{0}$ correction shown in a) and b) applied. Isosurfaces show the 95 \% probability boundary for all plots (i.e. there is a 95 \% probability the electron/hole will be found inside the plotted surfaces).}
	\label{fig:Toy_model}
\end{figure}

We plot DFT energies versus $G_{0}W_{0}$ energies for valence and conduction bands in Figure~\ref{fig:Toy_model}a and b for CsI terminated toy model with spin-orbit coupling. Unlike in most situations, corrections to DFT-level energies do not form straight lines. This is because the halide perovskite and tetracene energy corrections are significantly different in magnitude (as tetracene's electronic states are more localised). Furthermore, there are some mixed states including both part of the tetracene and halide perovskite wavefunction which carry intermediate energy corrections. To model tetracene's electronic states correctly at the interfaces, we fit a straight line to states where the DFT wavefunction is fully localised on tetracene, as shown on the figure. These fits allow for a comparison of tetracene's $G_{0}W_{0}$ level bandgap with that of bulk tetracene and isolated tetracene sheets (cut 1). We present results in supplementary information Table~\ref{tab:TcG0W0Eg}. Tetracene's $G_{0}W_{0}$ level bandgap is lowered at all interfaces, which is attributed to the halide perovskite stabilising single particle excited states in tetracene.

We show a schematic of energy levels in the toy model interfaces in Figure~\ref{fig:Toy_model}c. The only states close to the Fermi level are tetracene's valence band and the halide perovskite's conduction band. We define the energy difference between these states as $E_{Tc-P}$. This means we can explore exciton dissociation at the interface, but not full exciton transfer.

To carry out BSE (exciton) calculations, we applied scissor corrections to DFT energy levels. Scissor corrections shift DFT-level valence and conduction band states by constant gradients, as in the red lines on Figure~\ref{fig:Toy_model}a and b, and add a constant value to the DFT-level bandgap. Initially, we used the scissor corrections found from $G_{0}W_{0}$ fits i.e. correctly replicating tetracene's electronic states at the interface (as in Figure~\ref{fig:Toy_model}a and b). With these scissor corrections, for both CsI and PbI$_{2}$ terminated surfaces, we found tetracene's triplet state to be the lowest energy state at the interface. In calculations including spin-orbit coupling we identified `triplet' and `singlet' states as triply/singly degenerate dark/bright states with significant electron and hole contributions on tetracene. At both interfaces we found the singlet and triplet energies to be extremely comparable to those of a tetracene cut 1 layer in a vacuum (see supplementary information Table~\ref{tab:TcG0W0Eg}). For example, for CsI termination with spin-orbit coupling, we found the triplet states at $1.17$ eV, while for the same tetracene geometry in a vacuum (i.e. no halide perovskite present) a triplet energy of $1.11$ eV was calculated, in agreement within calculation error. This is indicative of tetracene's excitonic states being relatively unaffected by the presence of the halide perovskite.

We plot the average hole and electron positions of the lowest energy triplet and singlet states for CsI termination in Figure~\ref{fig:Toy_model}d and e respectively (with scissor correction applied which correctly reproduces tetracene's electronic states). Both states are found to be strongly localised to tetracene. Specifically, for both triplet and singlet there is less than a 3 \% probability of the electron or hole being within the halide perovskite. Importantly, there are lower energy charge transfer states than the singlet state in this model where the electron is fully localised on the halide perovskite. We present equivalent results for PbI$_{2}$ terminated surfaces in supplementary information~\ref{App:Exciton_PbI2}. Again we calculate well formed singlet and triplet states localised on tetracene only.

Figure\,\ref{fig:Toy_model} was in the case of $E_{Tc-P}$ being larger than tetracene's triplet energy. To explore excitonic states in the presence of lower energy states available, we then slightly altered the scissor correction, reducing $E_{Tc-P}$ and making charge transfer states (with the electron localised on the halide perovskite and the hole on the tetracene) lower in energy than tetracene's triplet energy (see supplementary information Table~\ref{tab:Scissors}). We obtained almost identical results: singlet and triplet states remain strongly localised on tetracene at the interface when there are lower energy states available. Furthermore, for charge transfer states, the electron/hole remained fully localised on the halide perovskite/tetracene. Our results therefore suggest one possible issue for exciton dissociation is the spatial separation of relevant states. 

Changing the scissor correction means tetracene's electronic states are no longer exactly reproduced fully at the interface. As a final confirmation to show that results with altered scissor corrections are valid, we replaced lead with tin to form an SnI$_{2}$ terminated surface. Following geometry relaxations, we carried out the same $G_{0}W_{0}$ and BSE calculations. We obtained charge transfer states lower in energy than tetracene's triplet energy for the physically correct scissor shift. Importantly, our results were found to be similar to those for the PbI$_{2}$ discussed in supplementary information~\ref{App:Exciton_PbI2}, with triplet states remaining localised on tetracene. This confirms that results calculated with approximate scissor corrections are valid.

We attempted to extend these calculations by further altering the scissor correction, reducing the halide perovskite's bandgap to less than tetracene's triplet energy. This required extreme scissor corrections and only resulted in charge transfer states (with electron localised on the halide perovskite only). No conclusions can be drawn from this as nothing resembling a singlet or triplet state was calculated. We anticipate this is due to the extreme scissor correction meaning tetracene's electronic states were not well reproduced.

In summary, we find tetracene's triplet and singlet states are strongly localised to tetracene at all interfaces considered. Furthermore, singlet and triplet energies are comparable to those for tetracene isolated in a vacuum for all interfaces. These combined results suggest that, for the optimal orientation of tetracene on the halide perovskite, tetracene's excitonic states still exist at the interface and are relatively unaffected by the presence of an inorganic semiconductor.

\section{\label{sec:Conclusion}Conclusion}

We have presented a study of the interface between singlet fission materials and halide perovskites. This study was motivated by 150 experiments screening for triplet transfer from a singlet fission material to a halide perovskite. In our model we found tetracene behaved in a broadly similar way to other organic/inorganic interfaces where the organic is a rod like structure: lone organic molecules orient parallel to the halide perovskite surface, while films orient with the long axis perpendicular to the surface. In all cases we found interface formation energies were much less favourable than for bulk tetracene and tetracene's valence band was higher energy than the halide perovskite's valence band, suggestive of tetracene being a good hole transporter. In general we observed only weak electronic interaction between these two materials. By using small models, we were able to calculate the excitonic states between tetracene and a toy halide perovskite-like structure. We found singlets and triplets remain localised on tetracene molecules at the interface and tetracene's excitonic energies were unaffected by the presence of the halide perovskite. Our results are indicative of only weak interaction between tetracene's excitonic states and an inorganic semiconductor (for optimal molecular arrangements at the interface). We suggest future work should focus on increasing electronic interaction at the interface by further exploring chemical bonding between the two materials, improving interface formation energies and increasing the probability of triplet transfer. Our work lays the ground for achieving triplet transfer from a singlet fission material to a halide perovskite.

\clearpage
\begin{acknowledgments}
A.R.B. acknowledges funding from a Winton Studentship, Oppenheimer Studentship and the Engineering and Physical Sciences Research Council (EPSRC) Doctoral Training Centre in Photovoltaics (CDT-PV). A.R.B. thanks all the support from the Yambo community forum, especially Daniele Varsano. A.R.B. acknowledged Matthew Klug, Rohit Prasanna, Thomas Feurer and Edoardo Ruggeri for fabrication of low-bandgap halide perovskites and copper indium gallium selenide thin films. S.D.S. acknowledges the Royal Society and Tata Group (UF150033) and the EPSRC (EP/R023980/1, EP/T02030X/1, EP/S030638/1). B.M. acknowledges support from the Gianna Angelopoulos Programme for Science, Technology, and Innovation and from the Winton Programme for the Physics of Sustainability. This work was performed using resources provided by the Cambridge Service for Data Driven Discovery (CSD3) operated by the University of Cambridge Research Computing Service (\url{www.csd3.cam.ac.uk}), provided by Dell EMC and Intel using Tier-2 funding from the Engineering and Physical Sciences Research Council (capital grant EP/P020259/1), and DiRAC funding from the Science and Technology Facilities Council (\url{www.dirac.ac.uk}). 

The data underlying this manuscript are available at [url to be added in proof].

S.D.S. is a Co-Founders of Swift Solar Inc. 
\end{acknowledgments}

\clearpage

\appendix

\section{List of experiments undertaken screening for triplet transfer}
\label{Screening_list}

Experiments break down into two types: evaporation and solution processing. Tetracene and DPH were purchased from Sigma at the highest purity available (> 99.99 \%, > 98 \% and > 95 \% respectively). All samples were fabricated on glass, which was cleaned in ultrasonic baths of acetone and then isopropanol prior to fabrication. All fabrication was carried out in nitrogen filled gloveboxes with oxygen levels of less than 10 ppm and water levels of less than 0.1 ppm. All samples were encapsulated immediately following fabrication, either by a two part or UV cured epoxy (Bluefixx). Halide perovskites marked `in situ' were fabricated by the author, with other samples supplied by Rohit Prasanna and Matthew Klug~\cite{Prasanna2017,Klug2016} (see section~\ref{Solution_screen} for fabrication details). In all cases no triplet transfer was observed. 

In general we found it was not straightforward to form a smooth interface between these materials, independent of the deposition method used. Instead, the two materials tended to separate, for example via the growth of pillars of the singlet fission material in evaporation (as in supplementary information Figure~\ref{fig:Schem_and_AFM}b). While some solution processing methods allowed for smooth bilayer outer surfaces, it was still unclear whether a smooth interface between the two materials had been achieved.

A wide parameter space for singlet fission material deposition was explored by using: 

\begin{enumerate}
	\item evaporation of the organic on the halide perovskite, with evaporation rates from 0.5 \AA$s^{-1}$ to 17 \AA$s^{-1}$ and deposition thicknesses from 7.5 nm to 150 nm
	\item spin coating the organic on the halide perovskite, with singlet fission materials dissolved in chlorobenzene (which halide perovskites are not soluble in), spinning speeds from 2000 to 4000 rotations per minute and sample annealing times from 0 to 30 minutes (prior to encapsulation)  
	\item drop casting the organic on the halide perovskite, with different solution concentrations deposited (though this resulted in extremely thick organic films in all cases, so was not significantly investigated)
	\item spin-coating the halide perovskite on singlet fission materials (noting this resulted in much of the singlet fission material being removed from the substrate)
	\item spin-coating the singlet fission material and halide perovskite at once to attempt to form heterojunctions.
\end{enumerate}

Full details of deposition conditions are now briefly listed.

\subsection{Evaporation}

All evaporations were carried out at a pressure of $\sim$ 1 $\times$ 10$^{-5}$ mbar or lower. Tetracene was evaporated at a temperature of 120$^{\circ}$C, DPH at110$^{\circ}$C, CBP at 190$^{\circ}$C and C$_{60}$ at 400$^{\circ}$C. Note for DPH evaporations thicknesses and evaporation rates are nominal as tetracene tooling factors were used. DPH is a heavier molecule so these should be regarded as upper bounds. 

\begin{enumerate}
	
\item 30 nm of tetracene was evaporated at a rate of 0.5 \AA s$^{-1}$ on FAPb$_{0.5}$Sn$_{0.5}$I$_{3}$ samples, with a small proportion of the lead and tin replaced with $x$=0 \%, 2 \%, 5 \% and 10 \% Ca, Mg, Sr, Zn, Co and Ni (where $x$ is defined as in Bowman and co-workers~\cite{Bowman2019}). Halide perovskite samples were $\sim$ 200 nm thick.

\item 30 nm of tetracene was evaporated at a rate of 0.5 \AA s$^{-1}$ on $x$=0 \%, 2 \% and 10 \% Mg, Sr and Co samples. Halide perovskite samples were $\sim$ 200 nm thick.

\item 60 nm of tetracene was evaporated at a rate of 1 \AA s$^{-1}$ on $x$=0 \%, 2 \% and 10 \% Co and Zn samples. Halide perovskite samples were $\sim$ 200 nm thick.

\item 60 nm of tetracene was evaporated at a rate of 0.7 \AA s$^{-1}$ on MA$_{y}$FA$_{1-y}$Sn$_{0.75}$Pb$_{0.25}$I$_{3}$ samples, with $y$=0, 0.2, 0.4, 0.6, 0.8 and 1. Halide perovskite samples were a range of thicknesses around 300 nm. 
\item 150 nm of DPH was evaporated at a rate of 0.5 \AA s$^{-1}$ on $x$=0 \%, 2 \%, 5 \% and 10 \% Zn samples. Halide perovskite samples were $\sim$ 200 nm thick.

\item 20/150/150 nm of DPH was evaporated at a rate of 0.5/0.8/7.0 \AA s$^{-1}$ on $\sim$ 200 nm thick Cs$_{z}$FA$_{1-z}$Sn$_{0.5}$Pb$_{0.5}$I$_{3}$ with $z$=0, 0.05, 0.1, 0.15 and 0.2, for $x$=0 \% and 5 \% Zn.

\item  150 nm of DPH was evaporated at a rate of 6.5/0.5/2.5/17 \AA s$^{-1}$ on $\sim$ 100 nm thick Cs$_{z}$FA$_{1-z}$Sn$_{0.75}$Pb$_{0.25}$I$_{3}$, $z$= 0, 0.05, 0.1, 0.15, 0.2, MA$_{y}$FA$_{1-y}$Sn$_{0.75}$Pb$_{0.25}$I$_{3}$, $y$=0.4, 0.8, 1.0 and in situ fabricated FA$_{0.75}$Cs$_{0.25}$Sn$_{0.75}$Pb$_{0.25}$I$_{3}$. 

\item 5.2 nm of C$_{60}$ was evaporated on MA$_{y}$FA$_{1-y}$Sn$_{0.75}$Pb$_{0.25}$I$_{3}$ with $y$=0, 0.8 and 1 at a rate of 0.4 \AA s$^{-1}$ (as C$_{60}$ is known to separate triplets from singlet fission materials). On these samples, and equivalent samples samples without C$_{60}$ deposited, 5/50/500/500 nm of DPH/DPH/tetracene/DPH was evaporated at a rate of 15 \AA s$^{-1}$ (noting that higher evaporation rates had been observed to reduce the formation of pillars). Halide perovskites were $\sim$ 100 nm thick.

\item $x$=0 \% and $x$=5 \% Zn samples of $\sim$ 200 nm and $\sim$ 50-100 nm  thicknesses had either DPH, or DPH followed by CBP evaporated on them. For the former, 100 nm of DPH was evaporated at 1 \AA s$^{-1}$, while for the latter 7.5 nm DPH was evaporated at 1.2 \AA s$^{-1}$ followed by 100 nm CBP at 1 \AA s$^{-1}$. This was with the suggestion of transferring singlet excitons from a high-bandgap absorber (CBP) to a thin layer of singlet fission material (DPH). 

\item 7.5 nm of DPH was evaporated at a rate of 5 \AA s$^{-1}$ on $\sim$ 500 nm and $\sim$ 100 nm thickness FASn$_{0.75}$Pb$_{0.25}$I$_{3}$ and MASn$_{0.75}$Pb$_{0.25}$I$_{3}$ samples. On half of these samples (four of each were made) 200 nm of CBP was evaporated at a rate of 1.2 \AA s$^{-1}$, again with the suggestion of transferring singlet excitons from a high-bandgap absorber (CBP) to a thin layer of singlet fission material (DPH). 

\end{enumerate}

\subsection{Solution processing}
\label{Solution_screen}

All solution processing was carried out in nitrogen filled gloveboxes with < 10ppm O$_{2}$ and < 1ppm H$_{2}$O.

All `in situ' fabricated low-bandgap FA$_{0.75}$Cs$_{0.25}$Sn$_{0.75}$Pb$_{0.25}$I$_{3}$ halide perovskites samples were fabricated from stock solutions of PbI$_{2}$ (0.28 M), SnI$_{2}$ (0.83 M), SnF$_{2}$ (0.16 M), FAI (0.83 M) and CsI (0.28 M) in a 65:35 solution of DMF:DMSO (all Sigma). In all cases films were spin coated: 30 $\mu$L of the solution was deposited on a substrate which was spun at 4000 rotations per minute for 30 s, with a gas quench applied from 15 s. Samples were annealed at 100$^{\circ}$C for 15 minutes. Solution processed experiments undertaken are as follows:

\begin{enumerate}
	
\item A stock solution for in situ fabrication was diluted to 10 \% of normal concentration and low-bandgap halide perovskites were deposited (achieving very thin films). 0.03/0.07/0.18 M DPH in chlorobenzene was prepared at 25/50/80$^{\circ}$C and statically deposited on the halide perovskite. Four of each sample were made, half of which were spun at 1000 rotations per minute for 20 s (i.e. spin-coating and drop casting was carried out). One of each sample type was annealed at 100$^{\circ}$C for 15 minutes while the others were left to air dry.

\item Bulk heterojunctions were spin-coated in situ. A FA$_{0.75}$Cs$_{0.25}$Sn$_{0.75}$Pb$_{0.25}$I$_{3}$ precursor solution was split into four. The first was used as a control solution, the second had 0.04 M DPH dissolved into it (DPH's solubility limit in these solutions at room temperature), the third was diluted by 50 \% and then had 0.04 M DPH dissolved into it and the last was heated to 70$^{\circ}$C and had 0.13 M DPH dissolved into it. 
\end{enumerate}

\section{Other experimental methods}
\subsection{Atomic force microscopy}

Atomic force microscopy (AFM) was carried out using an Asylum Research MFP-3D atomic force microscope in non-contact AC mode. 0\textsuperscript{th} order flattening and 1\textsuperscript{st} order plane fits were applied to all data. All measurements and data processing were carried out on Asylum Research AFM Software version 15.

\subsection{Photoluminescence}

In photoluminescence measurements samples were excited by a continuous wave temperature controlled Thorlabs 405 nm laser. The emission was recorded using an Andor IDus DU420A silicon detector for lead-only samples or and an Andor IDus DU490A InGaAs detector.

\subsection{Time resolved photoluminescence}

Time‐resolved PL spectra were recorded using a gated intensified CCD camera (Andor iStar DH740 CCI‐010) connected to a calibrated grating spectrometer (Andor SR303i). A Ti:sapphire optical amplifier (1 kHz repetition rate, 90 fs pulse width) was used to generate narrow bandwidth photoexcitation (20 nm full‐width at half maximum) with a wavelength of 400 nm, via a custom‐built noncollinear optical parametric amplifier. 

\section{Computational details}
\label{sec:Comp_details}

We carried out geometry optimisations with the density functional theory (DFT) code {\sc CASTEP} \cite{Clark2005} with on-the-fly generated ultra-soft pseudopotentials. Spin-orbit coupling was not included in our geometry optimisations as we found it had a small effect relative to the additional computational effort: the lattice parameter of cubic CsPbI$_{3}$ only changed from $6.155$\AA~to $6.185$\AA~with the inclusion of spin-orbit coupling. We used a cutoff energy of $400$ eV in all geometry optimisations. Van der-Waals semi-empirical corrections were required for all calculations to correctly reproduce tetracene's geometry and electronic structure, as is discussed further in section~\ref{sec:Tetracene_alone}. For primitive tetracene and halide perovskite unit cells we used a Monkhorst-Pack $\vec{k}$-point grid of $5\times 5 \times 5$, while for larger cells (e.g. two repeating tetracene and three repeating halide perovskite units in the non-vacuum direction) we used a commensurate reduced number of $\vec{k}$-points in periodic directions (with a minimum of 2 $\vec{k}$-points being used in these directions), while in the non-periodic direction only the $\Gamma$ point was sampled. In all supercells with a vacuum, we converged the vacuum size to be large enough to not affect results. This corresponded to a vacuum at least 0.8 times the size of the unit cell for CsPbI$_{3}$ and at least the same length as the unit cell size  for tetracene. The same cutoff energies and $\vec{k}$-point grids were used for density of states (DOS) and projected-DOS (PDOS) calculations. For density of states calculations including spin-orbit coupling we used a cutoff energy of $500$eV, a Monkhorst-Pack $\vec{k}$-point grid of $5\times 5 \times 5$ for SCF calculations, a Monkhorst-Pack $\vec{k}$-point grid of $6\times 6 \times 6$ for spectral calculations and {\sc CASTEP}'s norm-conserving pseudopotentials. For calculations of a tetracene molecule on a halide perovskite surface we fixed the lattice parameters, while for thin-film interfaces lattice parameters were allowed to vary freely (discussed further in section~\ref{sec:Molec_int} and~\ref{sec:Full_int}).

We used the DFT code {\sc Quantum Espresso} and post-DFT code {\sc Yambo} to calculate electronic and excitonic states\cite{Giannozzi2009,Giannozzi2017,Marini2009}. In {\sc Quantum Espresso} we used $\vec{k}$-point grids of $6 \times 6 \times 6$ (so the $\Gamma$ and R points were both directly sampled), with a cutoff energy of $680$ eV ($50$ Ry). Norm-conserving Vanderblit pseudopotentials, taken from the Schlipf-Gygi norm-conserving pseudopotential library, were employed in these computations as they are optimised for subsequent {\sc Yambo} calculations\cite{Hamann2013,Schlipf2015,Scherpelz2016}. In all our {\sc Yambo} calculations, parameters were converged to give results to an accuracy of at least $0.05$ eV. To aid with {\sc Yambo} calculations at full interfaces, we sometimes ignored the non-local commutator (noted in text), which was found to affect calculations on tetracene and halide perovskite only minimally (changing energies by <$0.05$ eV), and for toy models we reduced the maximum size of reciprocal lattice vectors (corresponding to the cut-off energy used in DFT calculations) with respect to that in {\sc Quantum Espresso} calculations (which affected the accuracy of calculations to < 0.01 eV).

All our visualisations were carried out with a combination of {\sc c2x} and {\sc Vesta}\cite{Rutter2018,Momma2011}.

\section{Functional choice}
\label{Functional_choice}
Geometry optimisations of tetracene were carried out with several functionals to assess which best reproduced experimental lattice parameters. Results are summarised in Table~\ref{tab:lattreslts}. We note that for PBE and PBEsol functionals the lattice expanded well beyond what would be physically reasonable, and did not lead to a relaxed geometry, without a van der Waals correction.

\begin{table}[h]
	\centering
		\begin{tabular}{ c | c }
		Functional & Lattice parameter \\
		\hline
		\hline
		Experimental~\cite{Campbell1962} & $\vec{a}=(7.9,0.0,0.0)$\\
		(no temperature stated,& $\vec{b}=(0.39,6.02,0.0)$\\
		assumed room temperature) & $\vec{c}=(-5.33,-2.08,12.26)$\\
		\hline
		LDA & $\vec{a}=(7.33,0.0,0.0)$\\
		& $\vec{b}=(0.48,6.01,0.0)$\\
		& $\vec{c}=(-5.46,-2.19,11.93)$\\
		Summed difference with experiment & 1.24 \\
		\hline
		PBE + TS & $\vec{a}=(7.68,0.0,0.0)$\\
		& $\vec{b}=(0.45,6.02,0.0)$\\
		& $\vec{c}=(-5.36,-2.19,12.09)$\\
		Summed difference with experiment & 0.60 \\
		\hline
		PBE + G06~\cite{Grimme2006} & $\vec{a}=(7.35,0.0,0.0)$\\
		& $\vec{b}=(0.55,6.14,0.0)$\\
		& $\vec{c}=(-5.88,-2.40,11.91)$\\
		Summed difference with experiment & 2.05 \\
		\hline
		PBE + JCHS~\cite{Jurecka2007} & $\vec{a}=(6.79,0.0,0.0)$\\
		& $\vec{b}=(0.65,5.90,0.0)$\\
		& $\vec{c}=(-5.76,-2.68,11.36)$\\
		Summed difference with experiment & 3.41 \\
		\hline
		PBESOL + TS & $\vec{a}=(7.53,0.0,0.0)$\\
		& $\vec{b}=(0.47,6.08,0.0)$\\
		& $\vec{c}=(-5.57,-2.23,12.04)$\\
		Summed difference with experiment & 1.12 \\
			
	\end{tabular}
	\caption{\label{tab:lattreslts}Lattice parameters obtained from geometry optimisations of bulk tetracene with different exchange correlation functionals, alongside experimental result. $\vec{a}$, $\vec{b}$ and $\vec{c}$ are the lattice vectors in Cartesian coordinates, and all lengths are in \AA. PBE and PBESOL functionals did not lead to a relaxed geometries without Van der Waals corrections.}
\end{table}

\section{Bulk CsPbI$_{3}$}
\label{App:Bulk_perov}

\begin{figure}[h] 
    \includegraphics[width=0.8\textwidth]{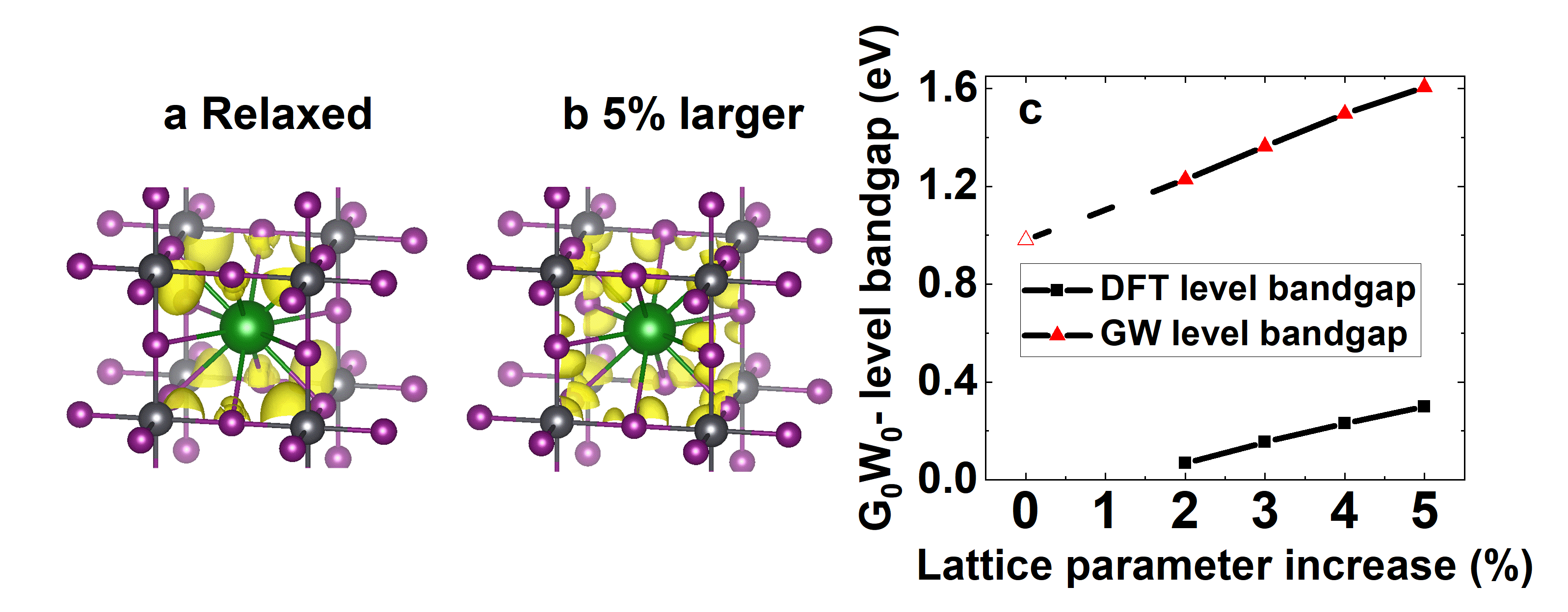}
    \caption{\label{fig:Bulk_perov_SI} a) and b) plot the valence band charge density (at the band edge) of cubic CsPbI$_{3}$ when fully relaxed, and with the lattice parameter increased by 5 \% (as marked on figure). The latter gives the experimental halide perovskite electronic structure with four regions of high charge density along a lattice parameter. As discussed in the main text, it was not possible to carry out $G_{0}W_{0}$ corrections on the relaxed halide perovskite. Instead, in c) $G_{0}W_{0}$ calculations for the lattice parameter increased between 2 \% and 5 \% are presented, which allows for an estimation of the relaxed $G_{0}W_{0}$ bandgap, as marked by the dashed line. All calculations include spin-orbit coupling.}
\end{figure}

Iodine and lead are the main contributors to the valence and conduction bands in experimental halide perovskites at room temperature~\cite{Haruyama2014}. However, in the fully relaxed cubic CsPbI$_{3}$, we found the nature of the valence and conduction bands is inverted at the DFT-level when including spin-orbit coupling (when using PBE with TS correction). We found increasing the halide perovskite's lattice parameter by 2 \% or more allows for correct band ordering at the DFT-level (with iodine/lead being the main contributor to the valence/conduction band), as plotted in Figure~\ref{fig:Bulk_perov_SI} a and b.

We carried out a one shot $G_{0}W_{0}$ correction to the cubic halide perovskite's band structure. While $G_{0}W_{0}$ corrections do not give halide perovskites' bandgaps with full accuracy, calculations beyond $G_{0}W_{0}$ are too computationally intensive to carry out when modelling interfaces~\cite{Filip2014,Leppert2019}. In order to have correct band ordering for calculations, we carried out $G_{0}W_{0}$ calculations on structures with the lattice parameter 2 \% to 5 \% larger than the relaxed structure. The DFT- and $G_{0}W_{0}$-level bandgaps for these cells (with spin-orbit coupling) are shown in Figure~\ref{fig:Bulk_perov_SI}c. From these results we extrapolated the approximate bulk bandgap of relaxed CsPbI$_{3}$ as 0.98 eV. This is close to the DFT-level bandgap without spin-orbit coupling of 1.09 eV, suggesting the halide perovskite's electronic structure without spin-orbit coupling approximately models results from post-DFT methods with spin-orbit coupling included. We use this to approximately model electronic states at interfaces in subsequent sections (and calculate approximate errors associated with this).

\section{Halide perovskite lattice sizes with a different number of repeating units in the vacuum direction}
\label{App:Perov_lat_size}

We present inter-atomic distances for geometry relaxed halide perovskite slabs with different numbers of periodic units in the vacuum direction in Table~\ref{tab:Perov_lat}. Specifically, we present the inter-atomic distance in the centre of the structure, in the vacuum direction, for both Cs-Cs and Pb-I distances, alongside the lattice parameter in a direction perpendicular to the vacuum. We note that the relaxed bulk cubic lattice parameter is $6.16$ \AA.

\begin{table}[h]
    \centering
    \begin{tabular}{m{3cm} | m{2cm} | m{2cm} | m{3cm}}
    Number of repeating units & Pb-I & Cs-Cs & Perpendicular lattice parameter\\
    \hline
    \hline
    \textit{CsI termination} & & & \\
    1 & $3.19$ & $4.89$ & $6.11$\\
    2 & $3.24$ & $5.19$ & $6.13$\\
    3 & $3.06$ & $5.59$ & $6.13$\\
    4 & $3.07$ & $5.79$ & $6.13$\\
    5 & $3.04$ & $5.99$ & $6.14$\\
    7 & $3.07$ & $6.13$ & $6.14$\\
    8 & $3.07$ & $6.15$ & $6.14$\\
    \hline
    \textit{PbI$_{2}$ termination} & & &\\
    1 & $3.06$ & N/A & $6.22$\\
    2 & $3.08$ & $5.79$ & $6.23$\\
    3 & $3.08$ & $5.93$ & $6.20$\\
    4 & $3.07$ & $6.10$ & $6.20$\\
    5 & $3.08$ & $6.10$ & $6.18$\\
    6 & $3.08$ & $6.17$ & $6.20$\\
    7 & $3.09$ & $6.18$ & $6.18$\\
    8 & $3.08$ & $6.14$ & $6.17$\\
    \end{tabular}
    \caption{\label{tab:Perov_lat}Inter-atomic distances in the centre of a halide perovskite slab, in the vacuum direction, and lattice parameters perpendicular to the vacuum, for both CsI and PbI$_{2}$ terminations. All values presented are in \AA.}
\end{table}

\section{Commensurate unit cells}
\label{App:Commens_uc}

\begin{figure}[h] 
    \includegraphics[width=0.3\textwidth]{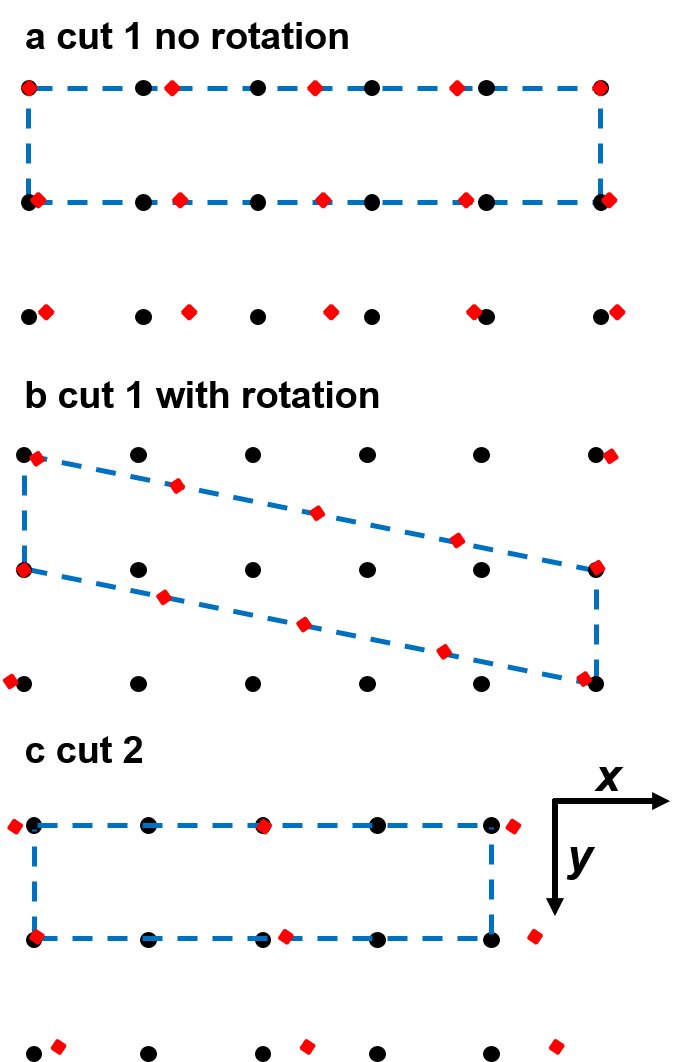}
    \caption{\label{fig:Commens_uc} Commensurate tetracene/CsPbI$_{3}$ unit cells. Black circles and red diamonds are halide perovskite and tetracene lattice points respectively. The three commensurate cells, `cut 1 no rotation', `cut 1 with rotation' and `cut 2' are presented in a), b) and c) respectively, with the dashed blue line representing one repeating unit.}
\end{figure}

We present the commensurate unit cells for halide perovskite and tetracene films in Figure~\ref{fig:Commens_uc}. The corresponding normal and shear strains for each arrangement are presented in Table~\ref{tab:Comms_strain}.

\begin{table}[h]
	\centering
	\begin{tabular}{l | c | c}
		& CsI termination & PbI$_{2}$ termination\\
		\hline
		\hline
		\textit{Cut 1 no rotation} & & \\
		Strain in \textit{x} direction (\%) & $-0.3$ & $0.9$ \\
		Strain in \textit{y} direction (\%) & $1.4$ & $2.6$ \\
		Shear strain (\%) & $7.5$ & $7.5$ \\
		\hline
		\textit{Cut 1 with rotation} & & \\
		Strain in \textit{x} direction (\%) & $1.7$ & $2.8$ \\
		Strain in \textit{y} direction (\%) & $1.4$ & $2.6$ \\
		Shear strain (\%) & $12.3$ & $12.3$ \\
		\hline
		\textit{Cut 2} & & \\
		Strain in \textit{x} direction (\%) & $1.4$ & $2.6$ \\
		Strain in \textit{y} direction (\%) & $-6.1$ & $-4.87$ \\
		Shear strain (\%) & $5.1$ & $5.1$ \\
	\end{tabular}
	\caption{\label{tab:Comms_strain}The normal and shear strain for each geometry presented in Figure~\ref{fig:Commens_uc}, relative to the relaxed halide perovskite slab in a vacuum.}
\end{table}

\section{G$_{0}$W$_{0}$ for different tetracene/halide perovskite electronic states and scissor corrections}
\label{App:G0W0corrections}
The effect of G$_{0}$W$_{0}$ corrections of different isolated tetracene and halide perovskite models, and small interfaces, are presented in Table~\ref{tab:G0W0corrections}. All models presented are in approximate agreement, with the tetracene valence band being increased in energy by $\sim$ $1$eV more than the halide perovskite valence band following a G$_{0}$W$_{0}$ correction, and the tetracene conduction band being increased by at least $1.5$eV more than the halide perovskite conduction band. 

For the CsI terminated toy model, the offset between the halide perovskite and tetracene's valence band is $0.97$eV without spin-orbit coupling and $0.78$eV with spin-orbit coupling. This suggests (alongside the results in Table~\ref{tab:G0W0corrections}) that the qualitative results drawn from PDOS calculations in sections~\ref{sec:Molec_int} and~\ref{sec:Full_int} (without spin-orbit coupling) would still be valid with the inclusion of spin-orbit coupling and subsequent G$_{0}$W$_{0}$ corrections, with tetracene valence band state being further increased in energy with respect to the halide perovskite valence band state.

We present tetracene's G$_{0}$W$_{0}$ bandgap, singlet and triplet energies for the toy interfaces, bulk tetracene and tetracene in a vacuum in Table~\ref{tab:TcG0W0Eg}. There are three models for tetracene in a vacuum -- geometry relaxed, and the same geometries as used at both PbI$_{2}$ and CsI terminated toy models. Finally, we present scissor shifts applied in the main text in Table~\ref{tab:Scissors}.

\begin{table}[h]
	\centering
	\begin{tabular}{m{6cm} | >{\centering\arraybackslash}m{3cm} | >{\centering\arraybackslash}m{3cm}}
		Model & $E_{Tc,VB,shift}-E_{P,VB,shift}$ (eV) & $E_{Tc,CB,shift}-E_{P,CB,shift}$ (eV)\\
		\hline 
		\hline 
		Bulk systems (no SO) & $1.53$ & $1.76$ \\
		Bulk systems (with SO) & $1.38$ & $1.88$ \\
		Bulk P and Tc in vacuum (no SO) & $0.83$ & $1.81$ \\
		Bulk P and Tc in vacuum (with SO) & $0.69$ & $1.92$ \\
		CsI terminated toy model (no SO) & $1.37$ & $1.20$ \\
		CsI terminated toy model (with SO) & $1.30$ & $1.40$ \\
		PbI$_{2}$ terminated toy model (no SO) & $1.37$ & $1.08$ \\
	\end{tabular}
	\caption{\label{tab:G0W0corrections}The energy difference in $G_{0}W_{0}$ corrections between tetracene and halide perovskite states, for different bulk and interface models. Here $E_{Tc,VB,shift}=E_{Tc,VB,G0W0}-E_{Tc,VB,DFT}$, the difference in energy between tetracene's (Tc) valence band energy at $G_{0}W_{0}$ and DFT levels. Other subscripts carry similar meanings, with P corresponding to halide perovskite and CB to conduction band. Lastly, SO corresponds to spin-orbit coupling.}
\end{table}

\begin{table}[h]
	\centering
	\begin{tabular}{l | c | c | c}
		Model & $E_{g,Tc,G0W0}$ (eV) & $E_{S,Tc}$ (eV) & $E_{T,Tc}$ (eV) \\
		\hline
		\hline
		Bulk tetracene & $2.62$ & $2.08$ & $1.20$ \\
		Tetracene in vacuum (relaxed) & $3.40$ & $2.04$ & $1.14$ \\
		Tetracene in vacuum (CsI toy geometry) & $3.28$ & $1.92$ & $1.11$ \\
		Tetracene in vacuum (PbI$_{2}$ toy geometry) & $3.30$ & $1.94$ & $1.12$ \\
		CsI terminated toy model (no SO) & $2.13$ & $1.91$ & $1.12$ \\
		CsI terminated toy model (with SO) & $2.64$ & $1.94$ & $1.17$ \\
		PbI$_{2}$ terminated toy model (no SO) & $2.24$ & $1.87$ & $1.08$ \\
	\end{tabular}
	\caption{\label{tab:TcG0W0Eg}Tetracene's $G_{0}W_{0}$ level bandgap ($E_{g,Tc,G0W0}$) and lowest singlet ($E_{S,Tc}$) and triplet ($E_{T,Tc}$) energies are presented for bulk tetracene, a layer of (cut 1) tetracene in a vacuum (both relaxed and for the geometry used in toy models) and for toy models with scissor corrections applied to correctly reproduce tetracene's electronic states. Here SO means spin-orbit coupling.}
\end{table}

\begin{table}[h]
	\centering
	\begin{tabular}{l | c | c | c}
		Model & $\Delta E_{g}$ (eV) & $M_{c}$ & $M_{v}$ \\
		\hline
		\hline
		\textit{CsI termination} &  &  &  \\
		Physically correct scissor correction & $1.88$ & $1.20$ & $1.23$ \\
		Alternative scissor correction & $1.30$ & $2.05$ & $1.23$ \\
		\hline
		\textit{PbI$_{2}$ termination} &  &  &  \\ 
		Physically correct scissor correction & $1.76$ & $1.11$ & $1.22$ \\
		Alternative scissor correction & $1.00$ & $2.75$ & $1.22$ \\
	\end{tabular}
	\caption{Scissor corrections applied to the toy interface models are presented. Here $\Delta E_{g}$ is the value added to the DFT-level bandgap and $M_{c}$ and $M_{v}$ are the gradient corrections applied to the DFT-level conduction and valence bands (for more details see {\sc YAMBO WIKI}~\cite{Yambo2021}). Physically correct scissor corrections are those found from fitting $G_{0}W_{0}$ calculations (c.f Figure~\ref{fig:Toy_model}) and alternative corrections are those applied to produce charge transfer states lower in energy than tetracene's lowest energy triplet.}
	\label{tab:Scissors}
\end{table}

\section{G$_{0}$W$_{0}$ calculations for PbI$_{2}$ terminated surfaces}
\label{App:Exciton_PbI2}

\begin{figure}[h]
	\centering
	\includegraphics[width=0.7\textwidth]{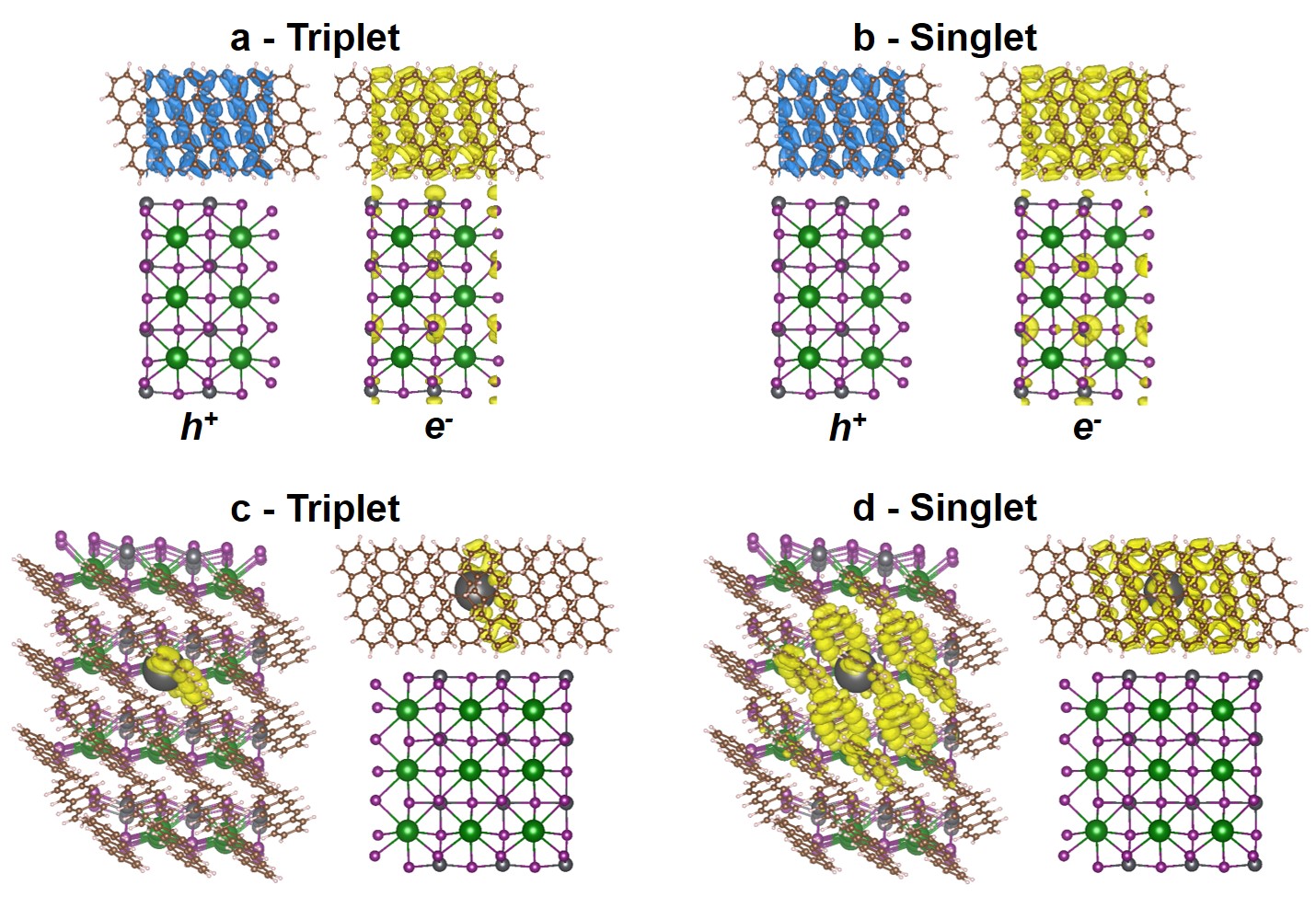}
	\caption{Average hole (h$^{+}$) and electron (e$^{-}$) charge densities for the lowest energy triplet and singlet excitons for the PbI$_{2}$ terminated toy model are plotted in a) and b), for the $G_{0}W_{0}$ correction which correctly reproduces tetracene's electronic states. The electron density of the same states with the hole (grey sphere) fixed on a tetracene molecule are plotted in c) and d), for two different unit cell orientations in each case. Isosurface marks the 95 \% probability boundary in all plots.}
	\label{fig:PbI2toymodel}
\end{figure}

In Figure~\ref{fig:PbI2toymodel}a and b we plot average exciton electron and hole position for PbI$_{2}$ termination  (for lowest energy triplet and singlet states). Here the average electron wavefunctions are found to have a small contribution within the halide perovskite (noting isosurfaces are only on tetracene up to the 88 \% probability surface). However, within our simulations tetracene makes up only a small region of the simulation cell and therefore there are contributions to the average electron position both from the hole being in vacuum and on the halide perovskite. To this end, we plot the electron charge density in Figure~\ref{fig:PbI2toymodel}c and d for the hole (grey sphere) fixed on a tetracene molecule. These plots reveal that when the hole is fixed on a tetracene molecule, the electron also is. We observed this for the hole being fixed at many different locations within tetracene. These results again demonstrate that tetracene's excitonic states are strongly localised, even at an interface. This is also suggestive that electrons in the halide perovskite may transfer into tetracene and then become localised within the organic.

\clearpage
\section{Additional supplementary information figures}

\begin{figure}[h] 
	\centering    
	\includegraphics[width=0.6\textwidth]{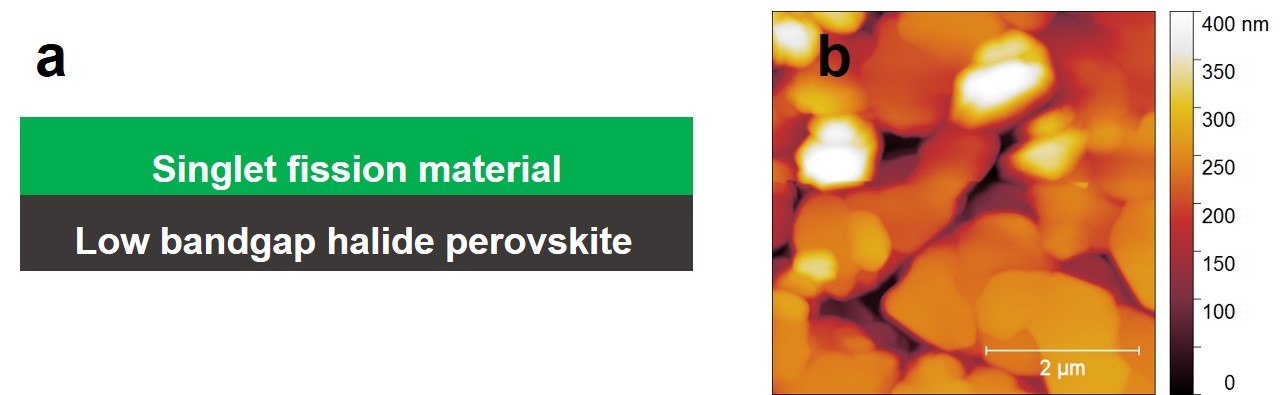}
	\caption[Schematic of a halide perovskite/singlet fission interface and atomic force microscopy and optical microscopy images of such interfaces.]{a) A schematic of a singlet fission/halide perovskite bilayer. b) an atomic force microscopy image of the surface of evaporated 1,6-Diphenyl-1,3,5-hexatriene (DPH).}
	\label{fig:Schem_and_AFM}
\end{figure}

\begin{figure}[h] 
	\centering    
	\includegraphics[width=0.8\textwidth]{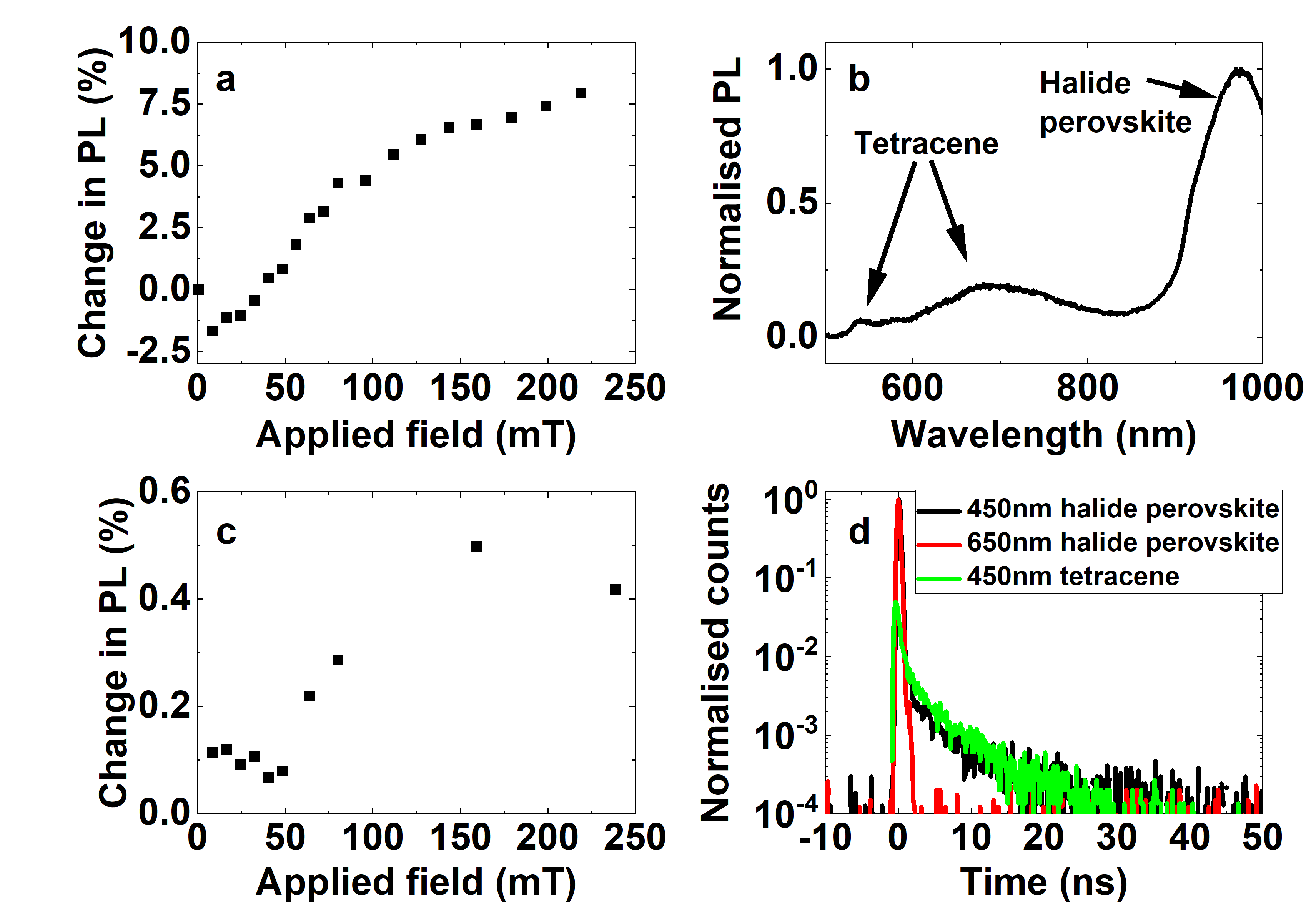}
	\caption[Typical results when screening bilayers for triplet transfer.]{a) The change in photoluminescence (PL) of tetracene to an applied magnetic field. The photoluminescence of a tetracene/FA$_{0.9}$Cs$_{0.1}$Pb$_{0.25}$Sn$_{0.75}$I$_{3}$ bilayer under 405 nm illumination is presented in b) (for tetracene evaporated on the halide perovskite). This image was taken with an InGaAs camera using a 500 nm long-pass filter, so the tetracene signal (in the 550-700 nm region) appears weak due to poor camera response in this region. The plot is cut at 1000 nm due to second order signal occurring at longer wavelengths. The halide perovskite PL peaks at $\sim$ 950 nm. c) The change in the halide perovskite's photoluminescence under 405 nm illumination following application of a magnetic field (where a 900 nm long-pass filter is used so only the halide perovskite's photoluminescence is observed). d) Time resolved photoluminescence of halide perovskite when exciting at 450 nm and 650 nm is presented, alongside suitably scaled TRPL from tetracene in the same bilayer under 450 nm illumination. The halide perovskite's photoluminescence is short-lived due to hole transfer to tetracene.}
	\label{fig:Tetracene_ex_results}
\end{figure}

\begin{figure}[h] 
\includegraphics[width=0.7\textwidth]{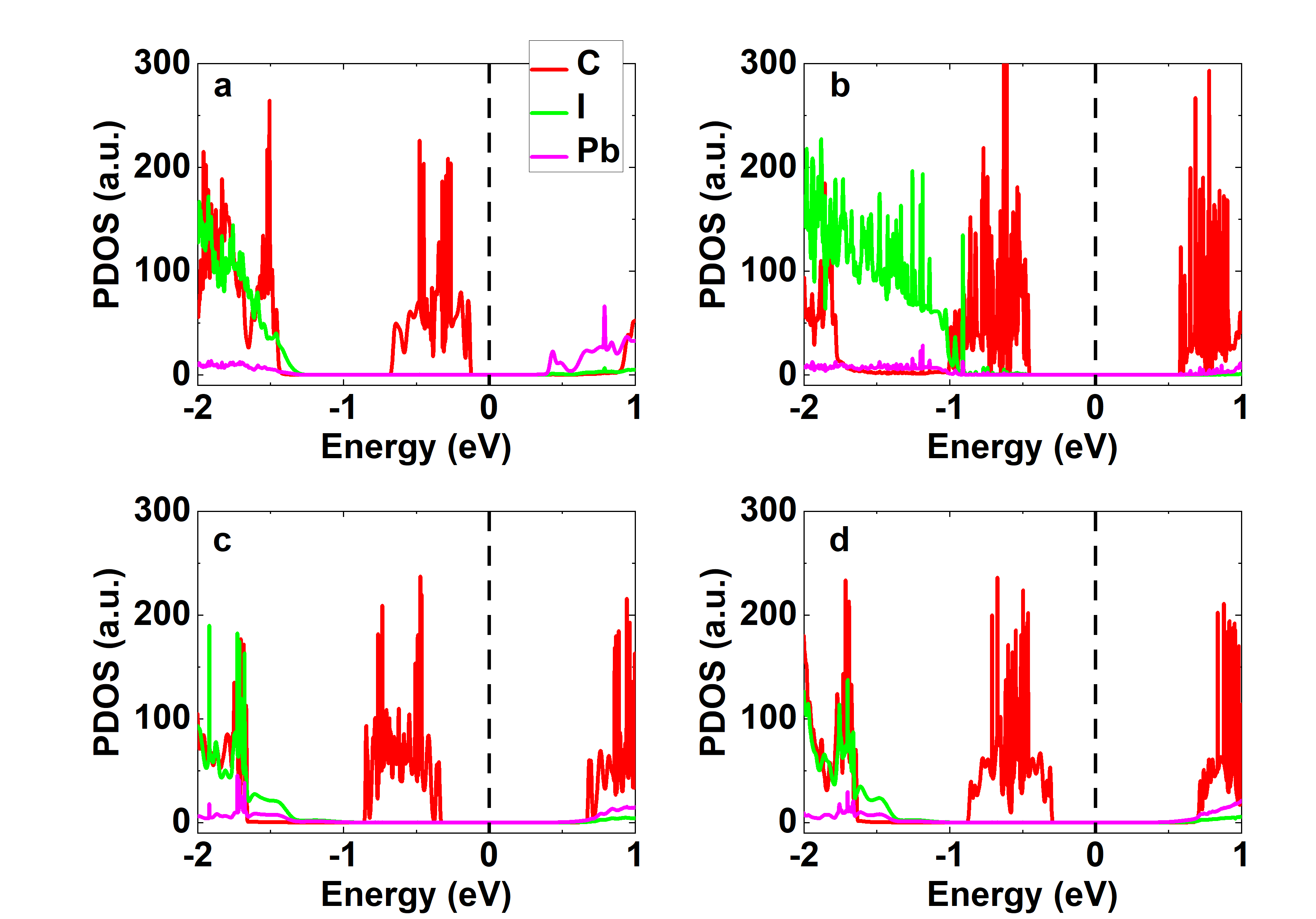}
\caption{\label{fig:OtherPDOS} PDOS, without spin-orbit coupling, are presented for relaxed interfaces of: cut 1 with rotation, PbI$_{2}$ termination; cut 2 PbI$_{2}$ termination; cut 1 no rotation, CsI termination; and cut 1 with rotation, CsI termination, in a), b), c) and d) respectively. Dashed vertical lines mark the Fermi level.}
\end{figure}

\begin{figure}[h] 
\includegraphics[width=0.7\textwidth]{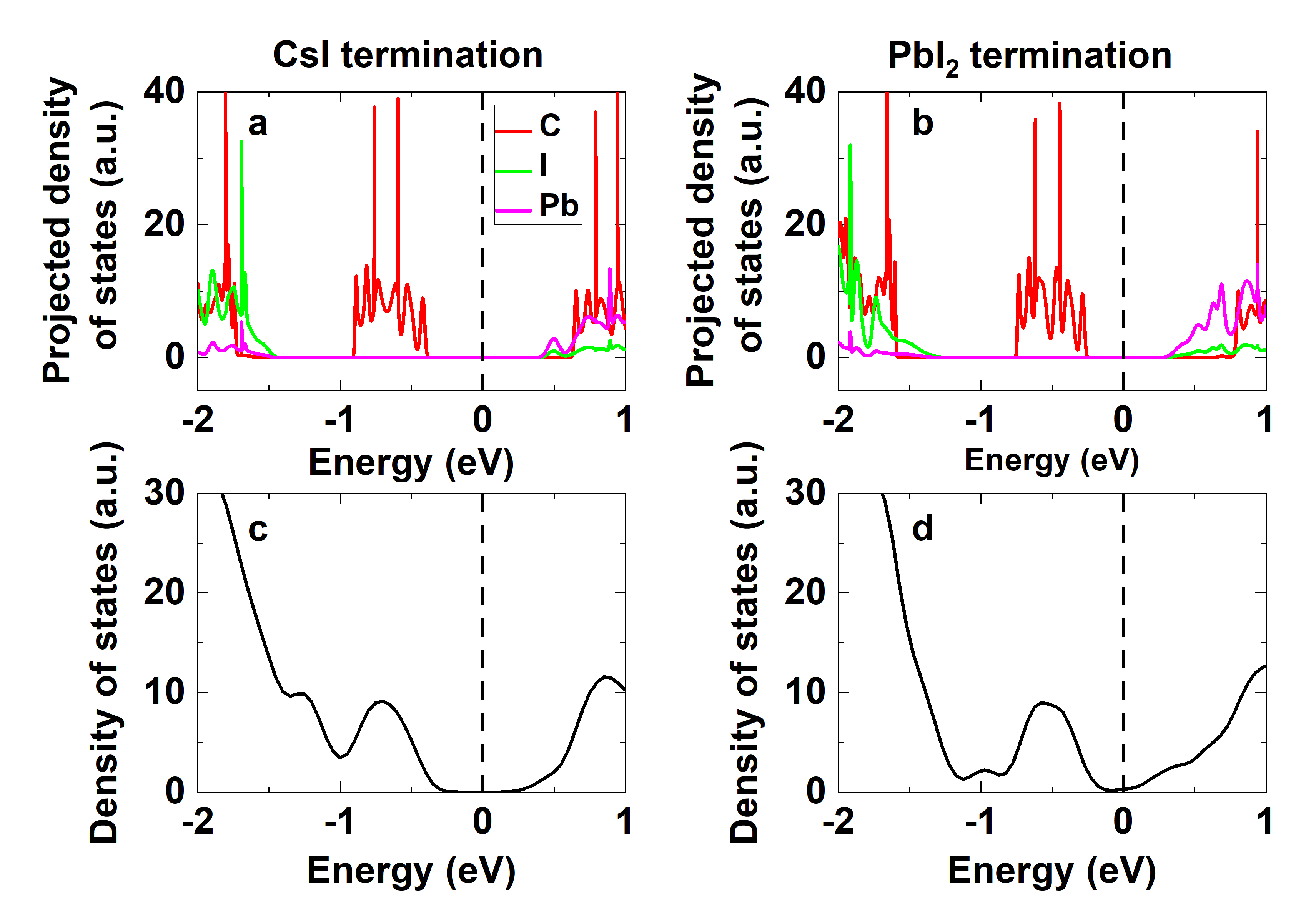}
\caption{\label{fig:ToyPDOS} a) and b) present the PDOS (without spin-orbit coupling) for CsI and PbI$_{2}$ terminated toy models. The density of states with spin-orbit coupling for the same models are shown in c) and d). In all plots the dashed line corresponds to the highest occupied level. Different smearing parameters have been used in plots with and without spin-orbit coupling.}
\end{figure}

\clearpage
\bibliography{library_manual}

\end{document}